\documentclass[fleqn,usenatbib]{mnras}

\usepackage{newtxtext,newtxmath}

\usepackage[T1]{fontenc}

\DeclareRobustCommand{\VAN}[3]{#2}
\let\VANthebibliography\thebibliography
\def\thebibliography{\DeclareRobustCommand{\VAN}[3]{##3}\VANthebibliography}

\usepackage{graphicx}	
\usepackage{amsmath}	
\usepackage{color,soul}
\newcommand{\teff}{$T_{\rm eff}$} 
\newcommand{\logg}{$\log g$} 
\newcommand{\gaia}{{\it Gaia}} 
\newcommand{\kms}{km~s$^{-1}$}
\newcommand{\ds}{$\Delta s$}
\newcommand{\dv}{$\Delta v$}
\newcommand{\vt}{$\xi_t$} 
\newcommand{\fei}{Fe\,{\sc i}}
\newcommand{\feii}{Fe\,{\sc ii}}
\newcommand{\bprp}{$BP-RP$} 
\newcommand{\msun}{$M_\odot$}
\usepackage{xcolor}
\usepackage{CJK}
\usepackage{orcidlink}
\usepackage[modulo]{lineno}

\usepackage{xcolor}
\usepackage{ulem}

\title[C3PO. I. Sample Selection and First Results]{C3PO: Towards a complete census of co-moving pairs of stars. I. High precision stellar parameters for 250 stars\thanks{This paper includes data gathered with the 6.5 m Magellan Telescopes located at Las Campanas Observatory, Chile. Some of the data presented herein were obtained at the W.\ M.\ Keck Observatory, which is operated as a scientific partnership among the California Institute of Technology, the University of California and the National Aeronautics and Space Administration. The Observatory was made possible by the generous financial support of the W.\ M.\ Keck Foundation. Based on observations collected at the European Southern Observatory under ESO programme 108.22EC.001.}}

\author[D. Yong et al.]
{David Yong\orcidlink{0000-0002-6502-1406},$^{1,2}$\thanks{E-mail: david.yong@anu.edu.au}
Fan Liu (刘凡)\orcidlink{0000-0003-4794-6074},$^{2,3,4}$ 
Yuan-Sen Ting (丁源森)\orcidlink{0000-0001-5082-9536},$^{1,2,5,6}$ 
Meridith Joyce\orcidlink{0000-0002-8717-127X},$^{2,7,8}$ \newauthor
Bertram Bitsch,$^{9}$  
Fei Dai\orcidlink{0000-0002-8958-0683},$^{10,11,12}$ 
Aaron Dotter,$^{13}$ 
Amanda I. Karakas\orcidlink{0000-0002-3625-6951}$^{2,3}$ and 
Michael T.\ Murphy\orcidlink{0000-0002-7040-5498}$^4$ 
\\
$^{1}$ Research School of Astronomy and Astrophysics, Australian National University, Canberra, ACT 0200, Australia\\
$^{2}$ ARC Centre of Excellence for Astrophysics in Three Dimensions (ASTRO-3D), Australia\\
$^{3}$ School of Physics and Astronomy, Monash University, Melbourne, VIC 3800, Australia \\
$^{4}$ Centre for Astrophysics and Supercomputing, Swinburne University of Technology, Hawthorn, VIC 3122, Australia \\
$^{5}$ School of Computing, Australian National University, Acton ACT 2601, Australia \\
$^{6}$ Department of Astronomy, The Ohio State University, Columbus, USA \\
$^{7}$ Konkoly Observatory, Research Centre for Astronomy and Earth Sciences, H-1121 Budapest Konkoly Th. M. \'ut 15-17., Hungary; meridith.joyce@csfk.org \\
$^{8}$ CSFK, MTA Centre of Excellence, Budapest, Konkoly Thege Mikl\'os \'ut 15-17., H-1121, Hungary;\\
$^{9}$ Max-Planck-Institut f\"ur Astronomie, Königstuhl 17, 69117, Heidelberg, Germany \\
$^{10}$ Division of Geological and Planetary Sciences, 1200 E California Blvd, Pasadena, CA, 91125, USA \\
$^{11}$ Department of Astronomy, California Institute of Technology, Pasadena, CA 91125, USA \\
$^{12}$ NASA Sagan Fellow \\
$^{13}$ Department of Physics and Astronomy, Dartmouth College, 6127 Wilder Laboratory, Hanover, NH 03755, USA}

\date{Accepted XXX. Received YYY; in original form ZZZ}

\pubyear{2015}

\begin{document}
\begin{CJK*}{UTF8}{gbsn} 
\label{firstpage}
\pagerange{\pageref{firstpage}--\pageref{lastpage}}
\maketitle

\begin{abstract}
We conduct a line-by-line differential analysis of a sample of 125 co-moving pairs of stars (dwarfs and subgiants near solar metallicity). We obtain high precision stellar parameters with average uncertainties in effective temperature, surface gravity and metallicity of 16.5~K, 0.033~dex and 0.014~dex, respectively. We classify the co-moving pairs of stars into two groups, chemically homogeneous (conatal; |$\Delta$[Fe/H]| $\le$ 0.04 dex) and inhomogeneous (non-conatal), and examine the fraction of chemically homogeneous pairs as a function of separation and effective temperature. The four main conclusions from this study are: (1) A spatial separation of \ds~=~10$^6$~AU is an approximate boundary between homogeneous and inhomogeneous pairs of stars, and we restrict our conclusions to only consider the 91 pairs with \ds~$\le$~10$^6$~AU; (2) There is no trend between velocity separation and the fraction of chemically homogeneous pairs in the range \dv~$\le$~4~\kms; (3) We confirm that the fraction of chemically inhomogeneous pairs increases with increasing \teff\ and the trend matches a toy model of that expected from planet ingestion; (4) Atomic diffusion is not the main cause of the chemical inhomogeneity. A major outcome from this study is a sample of 56 bright co-moving pairs of stars with chemical abundance differences $\leq$~0.02~dex (5\%) which is a level of chemical homogeneity comparable to that of the Hyades open cluster. These important objects can be used, in conjunction with star clusters and the \gaia\ ``benchmark'' stars, to calibrate stellar abundances from large-scale spectroscopic surveys. 

\end{abstract}

\end{CJK*}




\section{Introduction}

Stars born in the same gas cloud are remarkable laboratories to study stellar and Galactic astrophysics because such objects share the same age and chemical composition. Therefore, any differences in chemical composition between two stars that formed together could arise due to (i) limitations and/or systematics in the analysis and/or (ii) astrophysical processes. 

Regarding the former, binary stars are incredibly powerful objects to aid in the calibration and validation of stellar chemical compositions from large-scale spectroscopic surveys such as Gaia-ESO \citep{Gilmore:2012aa}, GALAH \citep{DeSilva:2015aa}, APOGEE \citep{Majewski:2017aa}, SDSS-V \citep{Kollmeier:2017aa} and 4MOST \citep{deJong:2019aa}. In this context, calibrations of stellar chemical compositions have relied upon stars in clusters and the 34 \gaia\ FGK ``benchmark'' stars \citep{Jofre:2014aa,Jofre:2015aa,Jofre:2017aa,Heiter:2015aa}. There are a number of significant advantages for using binary stars\footnote{Here and throughout we are referring to well-separated binary stars that have not interacted and whose evolution can be treated as single-star evolution from a modeling perspective.} to calibrate and/or validate stellar chemical compositions from spectroscopic surveys: (1) they are much more abundant than star clusters and the 34 \gaia\ FGK ``benchmark'' stars; (2) they cover a larger range of parameter space (temperature, mass, metallicity, age, location etc.); (3) they populate the parameter space more densely. Any new avenue to improve the calibration of stellar chemical abundances from large surveys would be profoundly important. 

Various astrophysical processes can produce differences in the chemical composition between two stars in a binary system. Those chemical abundance differences, however, can be subtle and difficult to detect. These processes include: 
\begin{enumerate}
    \item Exoplanet formation and/or engulfment. If one of the stars in the binary system has engulfed a planet, this will imprint a predictable chemical abundance signature onto the host star \citep{Chambers:2010aa}, and may therefore induce differences in abundances within a binary system \citep{tuccimaia:2014aa,ramirez:2015aa,Saffe:2017aa,Oh:2018aa,Liu:2018aa,Ramirez:2019aa, Liu:2021aa,Spina:2021aa}. Similarly, the formation of terrestrial planets may ``remove'' refractory element material from the host star (e.g., \citealt{Melendez:2009ab,Bitsch:2018aa}) leading to an apparent depletion of such elements within a binary system. 
    \item Stellar astrophysics. Atomic diffusion is a generic term used to describe a variety of mixing processes in the atmospheres of stars that can affect the apparent chemical composition of the star as a function of stellar evolution, i.e., stellar age and mass \citep[e.g.,][]{Korn:2007aa,Nordlander:2012aa,Dotter:2017aa}. For some conatal systems such as open clusters and binary star systems, atomic diffusion can induce small but detectable abundance differences  \citep{Souto:2018aa,Souto:2019aa,Liu:2019aa}. 
    \item Star formation and the interstellar medium. The chemical homogeneity of gas within a star forming cluster depends upon how the interstellar medium operates and mixes gas \citep{Feng:2014aa,Krumholz:2018aa}. By studying binary star systems with different spatial and velocity separations, we can essentially study the chemical homogeneity of the natal gas clouds at different spatial separations \citep{Kamdar:2019aa}. 
    \item Dust evolution and accretion. \citet{Hoppe:2020aa} examined how the growth and drift of dust particles in the protoplanetary disc can influence the abundances of stars. Since the pebbles drift faster than the gas (e.g., \citealt{Brauer:2008aa}), the heavy elements are accreted before the main gas leading to an enrichment of the star.  
\end{enumerate} 

High precision chemical abundances for a large sample of binary stars can, in principle, allow us to distinguish between these different astrophysical processes. In particular, pairs of stars with a range of differences in effective temperature, surface gravity, spatial separation and 3D velocity separation are needed to probe the above-mentioned processes. Motivated by these science questions, obtaining and analysing such a sample is the goal of this work. 

While the importance of binary stars has long-been recognised, the number of known binary stars that are sufficiently bright to enable high-precision chemical abundance analysis is small. At the time we embarked upon this study, to our knowledge only $\sim$10 binary star systems had high-precision (relative abundance errors of the order $\sim$0.01 dex; $\sim$2\%) chemical abundance examinations (see \citealt{Behmard:2023aa} and references therein). Those small numbers were due to (1) the lack of known binary stars and (2) random target selection in spectroscopic surveys; most surveys did not target both stars in a binary system nor do they routinely deliver (relative) abundance precision at the $\sim$0.01 dex level which was necessary to detect stellar chemical signatures of terrestrial planet formation. We also note that the planet engulfment hypothesis remains somewhat contentious and that subtle variations in abundance ratios [X/Fe] with stellar age can affect the interpretation \citep{Nissen:2015aa}. Therefore, conatal stars will enable a more robust identification of planet engulfment signatures.  
 
Before \gaia\ \citep{gaia}, it was a major challenge to (1) reliably identify binaries with wide separations and (2) distinguish binaries from chance alignments of stars at different distances. Using \gaia\ EDR3, \citet{elbadry:2021aa} provided a sample of 1 million binary star systems and the vast majority have sufficiently large separations and (presumably) orbital periods such that they do not interact and most ``can in some sense be viewed as clusters of two: the components formed from the same gas cloud and have orbited one another ever since''. For simplicity, we will refer to both the bound binary and unbound co-moving systems simply as ``co-moving pairs'' of stars. 

While \citet{elbadry:2021aa} probed binary stars with separations up to 1 pc, the simulations by \citet{Kamdar:2019aa} predicted that co-moving pairs of stars with significantly larger spatial separations (up to 20 pc; $\sim$4 $\times$ 10$^6$ AU) and 3D velocity separations < 1.5 \kms\ are conatal, i.e., born from the same gas cloud. If correct, then this would greatly increase the population of co-moving stars for high-resolution spectroscopic studies. Indeed our pilot study of 31 co-moving pairs \citep{Nelson:2021aa} confirmed that 73 $\pm$ 22\% of widely separated (10$^5$ - 10$^7$ AU) co-moving pairs of stars with 3D velocity differences $<$ 2 \kms\ are chemically homogeneous and thus presumably conatal. 

Motivated by the 1 million binary star sample of \citet{elbadry:2021aa}, we generated our own catalogue of co-moving pairs of stars with spatial separations as large as 30 pc following the approach of \citet{Nelson:2021aa}. Our expectation was that such a catalogue would include a significant number of sufficiently bright conatal co-moving pairs of stars from which we can obtain high-resolution, high signal-to-noise ratio spectra and thereby deliver high-precision chemical abundance measurements. 

Using the differential analysis techniques pioneered by \citet{Melendez:2009ab} and refined by \citet{Liu:2014aa} and \citet{ramirez:2014aa}, our goal is to obtain high precision relative chemical abundance measurements of a large sample of co-moving pairs of stars with internal errors as low as 0.01 dex ($\sim$2\%); a factor of five improvement over traditional techniques \citep{Nissen:2018aa}. 
Such high precision relative chemical abundance measurements are beyond the limit of any ongoing or planned large-scale spectroscopic surveys, yet essential for the main aims of this study. Those goals are to detect the subtle signatures of chemical inhomogeneity in the star forming clouds which could then be attributed to atomic diffusion, planet engulfment and/or other processes.

The goal of this paper is to introduce the sample selection, observations, analysis and first results of C3PO; towards a {\bf C}omplete {\bf C}ensus of {\bf C}o-moving {\bf P}airs {\bf O}f stars. The paper is arranged as follows. In Section 2 we describe the sample selection, observations and data reduction. In Section 3 we present the analysis focusing on the spectroscopically-determined stellar parameters and metallicity, [Fe/H] (other elements will be presented in a future work). Section 4 includes our results. Our discussion and conclusions are given in Sections 5 and 6, respectively. 

\section{Sample Selection, Observations and Data Reduction}

The sample was selected in the following way. Using data from \gaia\ EDR3 \citep{gaiaedr3}, spatial separations and 3D velocity separations were computed for the co-moving pairs using the same approach as in \citet{Nelson:2021aa}. We then applied the following criteria. 

\begin{enumerate}
    \item Spatial separation \ds\ $<$ 30 pc ($\sim$ 10$^{6.8}$ AU). The cutoff value was chosen to best balance the number of targets available for observation while recognising that increasing this limit would increase contamination from chance alignments and non-conatal pairs.  
    \item 3D velocity separation, \dv\ $<$ 2.0 \kms. As above, the cutoff value was chosen to best balance the number of targets while limiting contamination from non-conatal pairs. 
    \item 0.65 $\le$ (\bprp) $\le$ 1.15~mag. The blue cutoff value was selected to avoid fast rotating stars for which accurate equivalent widths, and therefore stellar parameters, are more difficult to measure with high precision. The red cutoff value was chosen to avoid cool stars for which the increased line blending makes accurate equivalent widths more difficult to obtain.  
    \item $| \Delta$~(\bprp)$| <$~0.15~mag. This criterion was included to ensure that for a given co-moving pair, both stars had sufficiently similar effective temperatures such that the differential analysis would yield small relative abundance uncertainties. 
    \item $| \Delta M_G| <$ 1 mag. While we wanted to ensure that the stars could have different evolutionary status to study atomic diffusion, we also needed to keep the surface gravities sufficiently close to enable high abundance precision. 
    \item G mag $<$ 10 mag. The cutoff was arbitrarily set to ensure that a sufficiently large sample size could be observed even when using 8-10m class telescopes. (Note that some of the objects in the pilot study by \citet{Nelson:2021aa} are fainter than G mag = 10.)  
    \item We generated a ``friends-of-friends'' search assuming a connecting threshold of 1pc in the 3D distances and omit all groups that have 5 or more members. Since the stars are nearby (90\% are within 200 pc), we assumed the uncertainties from the 3D Gaia position (including parallax) to be negligible. This criterion effectively removed most known open clusters and moving groups, allowing us to focus on true wide binaries that are dispersed in the Milky Way.
\end{enumerate}
There were 283 co-moving pairs of stars which simultaneously satisfied these criteria, the majority of which are dwarfs and subgiants. High-resolution spectroscopic observations of a subset of these targets were obtained using the Magellan Telescope (7 nights; 78 pairs), Keck Telescope (1 night; 23 pairs) and the European Southern Observatory's (ESO) Very Large Telescope (26.4 hours; 25 pairs). The remaining co-moving pairs were not observed or included objects with sufficiently large line broadening ($V_{\rm broad}$ $>$ $\sim$ 10 \kms) such that accurate equivalent widths, and thus stellar parameters, could not be obtained using the techniques described in the following section.

Observations taken with the Magellan Telescope made use of the Magellan Inamori Kyocera Echelle (MIKE) spectrograph \citep{Bernstein:2003aa} on 26-27 August 2021 and 11 November 2021. The slit width was set to 0\farcs5 which resulted in a spectral resolution of 50,000 and 40,000 in the blue (3350-5000~\AA) and red (4900-9300~\AA) CCDs, respectively. The CCD binning was set to 1 $\times$ 1 and exposure times adjusted to achieve signal-to-noise ratios of at least S/N = 200 per pixel near 5500~\AA. We also re-analysed the spectra from the pilot study of \citet{Nelson:2021aa} which were also taken using the Magellan Telescope and the MIKE spectrograph on 13-16 June 2019. Those observations have the same spectral resolution and wavelength coverage. As reported in \citet{Nelson:2021aa}, the median S/N per pixel for the blue and red CCDs were 121 and 185, respectively. We note that some of those co-moving pairs do not satisfy the selection criteria noted above: four pairs have 50 pc $\le$ \ds\ $\le$ 100 pc and two pairs have \dv\ = 3.05 and 4.4 \kms\ (those pairs were observed as a ``control'' sample). The spectra were reduced using the CarPy data reduction pipeline\footnote{\url{https://code.obs.carnegiescience.edu/mike}} described in \citet{Kelson:2003aa}. 

Keck Telescope observations were taken on 17 December 2021 and 16 January 2022. We used the High Resolution Echelle Spectrometer (HIRES; \citealt{Vogt:1994aa}) with the 0\farcs574 slit which resulted in a spectral resolution of 72,000 and a wavelength coverage of 4200 to 8500~\AA. The CCD binning was set to 1 $\times$ 1 and the exposure times were adjusted to achieve S/N $\ge$ 200 per pixel near 5500~\AA. The Keck-MAKEE pipeline\footnote{\url{https://sites.astro.caltech.edu/~tb/makee/}} was used for standard echelle spectra reduction. 

Observations with ESO's VLT were taken using the Ultraviolet and Visual Echelle Spectrograph (UVES; \citealt{Dekker:2000aa}) in service mode during Period 108 (which spanned 1 October 2021 through 31 March 2022). We used the 580nm setting, image slicer \#3, 1 $\times$ 1 CCD binning and the 0$\farcs$3 slit which resulted in a wavelength coverage of 4800 to 6800~\AA\ and spectral resolution of 110,000. Data reduction was performed using the ESO Reflex UVES pipeline v.6.1.6 \citep{Freudling:2013aa}. The exposure times were adjusted to achieve S/N $\ge$ 200 per pixel near 5500~\AA. 

\begin{figure}
	\includegraphics[width=.75\hsize,angle=90]{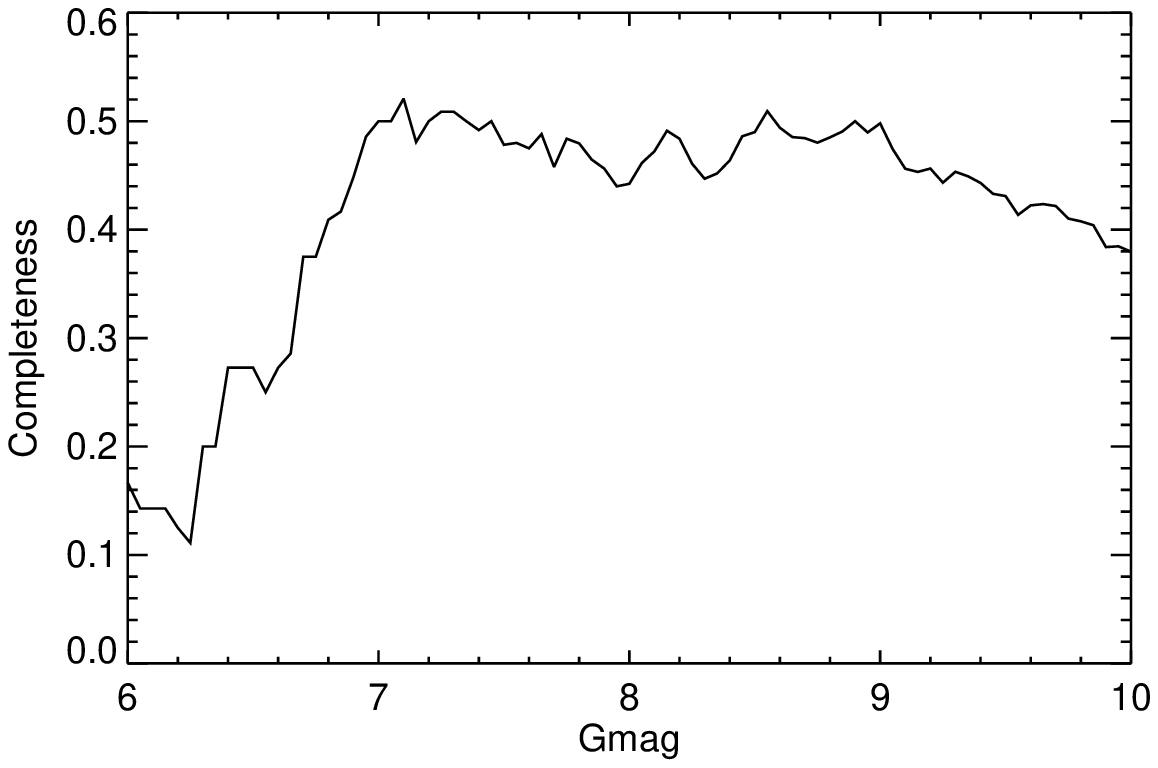}
    \caption{Completeness of observed sample as a function of $G$mag. Selection criteria are $\Delta s$ $<$ 30 pc, $\Delta v$ $<$ 2.0 km s$^{-1}$, 0.65 $\le$ ($BP-RP$) $\le$ 1.15~mag, $| \Delta$~($BP-RP$)$| <$~0.15~mag, $| \Delta M_G|~<$ 1 mag, G mag $<$ 10 mag and ``friends-of-friends'' restriction as described in Sec 2.}
    \label{fig:completeness}
\end{figure}

In total, we have observations of 125 co-moving pairs, i.e., 250 stars (see Table \ref{tab:obs}). To our knowledge this represents the largest sample of co-moving stars ever examined in a single high-precision differential analysis and some 44\% of the sample defined above. The completeness as a function of \gaia\ G mag is presented in Figure \ref{fig:completeness}. 

When necessary, multiple exposures were combined and individual orders were normalised using routines in IRAF\footnote{IRAF is distributed by the National Optical Astronomy Observatories, which are operated by the Association of Universities for Research in Astronomy, Inc., under cooperative agreement with the National Science Foundation.}. Orders were merged to create a single continuous spectrum per star. 

\begin{table*}
	\caption{Program stars and observing details.}
	\label{tab:obs}
	\begin{tabular}{llcclrcccccc}
	\hline
  Pair ID &
  Gaia\_ID\_ref &
  G\_ref &
  bp\_rp\_ref &
  Gaia\_ID\_obj &
  G\_obj &
  bp\_rp\_obj &
  $\Delta$ s ($\log$ AU) &
  $\Delta$ v (\kms) &
  Anom &
  Observed \\
\hline
   1 &     55780840513067392 &   8.373 &  0.783 &     55780840513308160 &   9.321 &  0.715 &   5.19 &   1.50 &  Y &  C,D \\ 
   2 &    627699888238838528 &   8.216 &  0.870 &    627699888238838272 &   7.458 &  0.742 &   4.29 &   0.51 &  N &    F \\ 
   3 &    692119656035933568 &   8.121 &  0.732 &    692120029700390912 &   8.144 &  0.738 &   4.81 &   0.75 &  N &    G \\ 
   4 &    704994524881597184 &   9.851 &  0.734 &    704994524881597056 &   9.855 &  0.733 &   4.90 &   0.78 &  N &    F \\ 
   5 &    736173925863826944 &   9.195 &  0.892 &    736174028943041920 &   9.206 &  0.884 &   4.40 &   0.80 &  Y &    G \\ 
   6 &    775037328283498624 &   9.443 &  0.786 &    741830466512452736 &   9.825 &  0.860 &   6.47 &   0.56 &  N &    G \\ 
   7 &    773252069293117696 &   8.973 &  0.786 &    773252069293612672 &   9.714 &  0.764 &   5.94 &   1.97 &  N &    G \\ 
   8 &    844865117036623232 &   7.556 &  0.802 &    844865117036622976 &   8.003 &  0.951 &   5.83 &   2.90 &  Y &    G \\ 
   9 &   1549927395024812672 &   8.225 &  0.741 &    844865117036623232 &   7.556 &  0.802 &   6.71 &   0.62 &  Y &    G \\ 
  10 &   1038148059924483328 &   8.701 &  0.775 &   1040472083909088128 &   9.317 &  0.906 &   6.19 &   1.68 &  Y &    G \\ 
	\hline
	\end{tabular}
	\\ 
	A = Magellan, June 2019; B = Magellan, 13 Aug 2021; C = Magellan, 26 Aug 2021; D = Magellan, 27 Aug 2021; E = Magellan, 11 Nov 2021; F = Keck, 20 Dec 2021; G = Keck, 16 Jan 2022; H = ESO, Period 108. \\
    (Recall that while the labels ``reference'' and ``object'' are interchangeable, we retain the distinction here to signify the manner in which the differential analysis was performed; see Sec 3 for details.) \\ 
	This table is published in its entirety in the electronic edition of the paper. A portion is shown here for guidance regarding its form and content.
\end{table*}

\section{Analysis}

Our analysis was conducted on a line-by-line differential manner similar to that described in \citet{Liu:2020aa,Liu:2021aa} but using a two-step process as described below. The primary advantage of conducting a differential analysis (for high quality spectra of stars with similar stellar parameters) is that the relative abundance errors can be as low as $\sim$0.01 dex (2\%). For a given co-moving pair, we arbitrarily refer to the two components as ``star A'' and ``star B'', i.e., the labels are interchangeable. In the first step, the stellar parameters for star A of a given co-moving pair were established with respect to the Sun. In the second step, the stellar parameters for star B were obtained relative to star A. We now describe the process but refer the reader to \citet{Melendez:2012aa}, \citet{Liu:2014aa,Liu:2018aa}, \citet{ramirez:2014aa} and \citet{Nissen:2018aa} for more details and discussion of the advantages and limitations of the technique. 

The line strengths (equivalent widths, EWs) were measured in each star for a set of lines taken from \citet{Liu:2014aa} and \citet{Melendez:2012aa}. The line list and EW measurements are presented in Table \ref{tab:ew}. We note that not all lines were measured in all stars and that typical uncertainties in the measured EWs are $<$ 1m\AA~due to the high S/N. The lines were assumed to have a Gaussian shape and we restricted the analysis to lines with EW $\le$ 150 m\AA, and we used the stellar line analysis program MOOG \citep{Sobeck:2011aa,Sneden:1973aa} and one dimensional local thermodynamic equilibrium (LTE) model atmospheres from \citet{Castelli:2003aa}. The effective temperature, \teff, was obtained by imposing excitation balance for \fei\ lines on a differential basis with respect to the reference star. The surface gravity, \logg, was determined by forcing ionization balance for \fei\ and \feii\ lines with respect to the reference star. The microturbulent velocity, \vt, was adjusted until the abundance differences exhibited no trend with the reduced EW ($\log$ EW/$\lambda$). When necessary, outliers ($>$ 3-$\sigma$) were removed and the process was repeated. In the first step, the Sun was the reference star and the stellar parameters were set as \teff\ = 5772 K, \logg\ = 4.44 dex\footnote{Here and throughout the paper, we use cgs units for \logg.}, \vt\ = 1.00 \kms, and [Fe/H] = 0.00 dex. The result of this first step were stellar parameters obtained in a differential manner with respect to the Sun. 

\begin{table}
        \centering
        \caption{Line list and equivalent width measurements.}
        \label{tab:ew}
        \begin{tabular}{lccrrr}
        \hline
        Wavelength &
        Species$^a$ &
        L.E.P.$^b$ &
        $\log gf$ &
        EW &
        EW \\
        (\AA) &
        &
        (eV) &
        &
        (m\AA) &
        (m\AA) \\
        \hline
         &        &        &           &  Pair 1 obj & Pair 1 ref \\
4445.47  &  26.0  &  0.09  &  $-$5.44  &  24.5  &  38.2 \\
4602.00  &  26.0  &  1.61  &  $-$3.15  &  63.2  &  74.2 \\
4630.12  &  26.0  &  2.28  &  $-$2.52  &  64.7  &  76.9 \\
4745.80  &  26.0  &  3.65  &  $-$1.27  &  69.6  &  81.6 \\
4779.44  &  26.0  &  3.42  &  $-$2.16  &  32.4  &  40.8 \\
        \hline
        \end{tabular}
        \\
                $^a$The digits to the left of the decimal point are the atomic number. The digit to the right of the decimal point is the ionization state ('0' = neutral, '1' = singly ionized). \\ 
                $^b$Lower excitation potential. \\
                This table is published in its entirety in the electronic
edition of the paper. A portion is shown here for guidance regarding its form
and content.
\end{table}

In the second step, the stellar parameters for star B (``object'' in Table \ref{tab:param}) were obtained using the same line-by-line differential approach described above. The principle difference in step two was that the reference star was star A of the co-moving pair, and the stellar parameters for star A (``reference'' in Table \ref{tab:param}) were obtained using the process described above. The stellar parameters\footnote{We emphasise, here and throughout, that we are reporting ``spectroscopically-determined stellar parameters''.} for the program stars are provided in Table \ref{tab:param}.  

\begin{table*}
	\caption{Stellar parameters.}
	\label{tab:param}
	\begin{tabular}{llccrlccccrc}
	\hline
  Pair ID &
  Gaia\_ID\_ref &
  \teff &
  \logg &
  [Fe/H] &
  Gaia\_ID\_obj &
  \teff &
  $\sigma$\teff &
  \logg &
  $\sigma$\logg &
  [Fe/H] &
  $\sigma$[Fe/H] \\ 
  &
  (ref) &
  (ref) &
  (ref) &
  &
  (obj) &
  (obj) &
  (obj) &
  (obj) &
  (obj) &
  (obj) \\ 
\hline
   1 &     55780840513067392 & 5986 &  3.970 &   0.082 &     55780840513308160 & 6164 &   10 &  4.240 &  0.024 &   0.014 &  0.008 \\ 
   2 &    627699888238838528 & 5579 &  4.540 &  -0.195 &    627699888238838272 & 6021 &   18 &  4.530 &  0.032 &  -0.220 &  0.015 \\ 
   3 &    692119656035933568 & 6003 &  4.560 &  -0.355 &    692120029700390912 & 6004 &   12 &  4.600 &  0.022 &  -0.333 &  0.009 \\ 
   4 &    704994524881597184 & 6034 &  4.490 &  -0.289 &    704994524881597056 & 6070 &   13 &  4.530 &  0.023 &  -0.291 &  0.008 \\ 
   5 &    736173925863826944 & 5589 &  4.520 &   0.159 &    736174028943041920 & 5615 &   10 &  4.530 &  0.017 &   0.124 &  0.007 \\ 
   6 &    775037328283498624 & 5836 &  4.470 &  -0.027 &    741830466512452736 & 5628 &   16 &  4.480 &  0.034 &  -0.027 &  0.015 \\ 
   7 &    773252069293117696 & 5979 &  4.080 &   0.215 &    773252069293612672 & 6036 &    8 &  4.360 &  0.021 &   0.219 &  0.008 \\ 
   8 &    844865117036623232 & 5900 &  4.450 &   0.249 &    844865117036622976 & 5620 &   29 &  4.510 &  0.044 &   0.086 &  0.021 \\ 
   9 &   1549927395024812672 & 6084 &  4.490 &   0.126 &    844865117036623232 & 5902 &   11 &  4.450 &  0.030 &   0.251 &  0.015 \\ 
  10 &   1038148059924483328 & 5992 &  4.570 &   0.068 &   1040472083909088128 & 5456 &   14 &  4.510 &  0.024 &  -0.216 &  0.011 \\ 
	\hline
	\end{tabular} \\ 
	This table is published in its entirety in the electronic edition of the paper. A portion is shown here for guidance regarding its form and content.
\end{table*}

\begin{figure}
	\includegraphics[width=.6\hsize,angle=90]{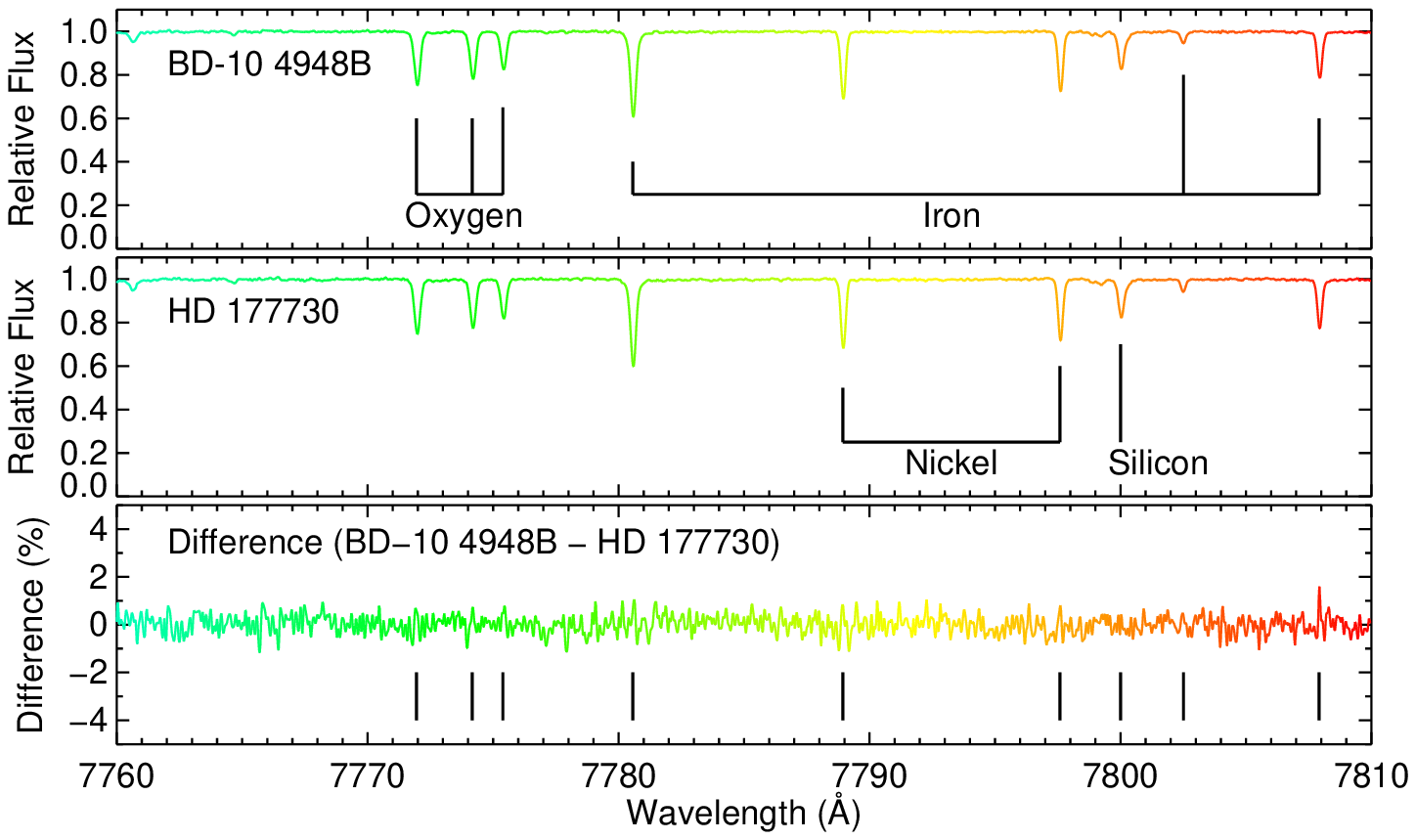}
    \caption{A portion of the spectra for the co-moving pair BD-10 4948B (upper) and HD 177730 (middle); Pair ID = 57. The spectrum in the lower panel is the difference between the two objects. Representative lines of O, Si, Fe and Ni are indicated. These two stars have essentially identical stellar parameters and the difference in the spectra is negligible. (A black and white version is included in the supplementary materials.)}
    \label{fig:spectra}
\end{figure}

The two stars in the co-moving pair, BD$-$10 4948B / HD 177730 (Pair ID = 57) have essentially identical stellar parameters; \teff\ = 5919/5923, \logg\ = 4.50/4.51 and [Fe/H] = $-$0.082/$-$0.074. In Figure \ref{fig:spectra}, we show a portion of the spectra for this co-moving pair which were observed using the MIKE spectrograph. In the region displayed in this figure, which includes lines of O, Si, Fe and Ni, the two spectra are indistinguishable as expected given their very similar parameters. This figure serves to demonstrate the power of the differential analysis technique when applied to high resolution, high S/N spectra. 

Recall that the sample were selected to have \teff\ similar to the Sun in order to minimise the relative errors resulting from a differential analysis. The hottest and coolest stars in our sample have \teff\ values of 6501 and 4929 K, respectively. The highest and lowest values of \logg\ are 4.60 and 3.84 dex, respectively. The most metal-poor object has [Fe/H] = $-$0.63 and the most metal-rich has [Fe/H] = +0.40 dex. We note that the majority of stars are dwarfs or early subgiants near solar metallicity. 

Following \citet{Liu:2014aa}, the uncertainties in the stellar parameters were obtained using the same approach as described in \citet{epstein:2010aa}. This procedure takes into account covariances between stellar parameters and the standard deviation of the differential abundances. We note the following uncertainties: for \teff\ the average uncertainty was 16.5 K with values ranging from 6.7 to 41.6K; for \logg\ the average uncertainty was 0.033 with values ranging from 0.012 to 0.090 dex; for metallicity, [Fe/H], the average uncertainty was 0.014 dex (i.e., 3\%) with values ranging from 0.006 to 0.026 dex. We examined the uncertainties in \teff, \logg, \vt\ and [Fe/H] as a function of these parameters. For all possible combinations, we note that there were no obvious nor significant trends. Additionally, we find that the magnitude of uncertainties in a given stellar parameter increase with the other uncertainties, i.e., as the uncertainty in \teff\ increases, so does the uncertainty in \logg.  

We emphasise that these are ``relative'' errors from our strictly differential line-by-line analysis and that these errors are considerably smaller than what is usually achieved in traditional analyses. For this study, it is the differential abundances, and the corresponding differential uncertainties, that matter when searching for chemical abundance differences within a given co-moving pair. For comparison, we note that the average errors from the detailed systematic study of $\sim$700 stars by \citet{bensby:2014aa} are $\sigma$ \teff\ = 63.5 K, $\sigma$ \logg\ = 0.095 dex and $\sigma$ [Fe/H] = 0.065 dex. That is, our ``differential'' errors are a factor of 3.8, 2.9 and 4.6, smaller for \teff, \logg\ and [Fe/H], respectively, when compared to the errors of \citet{bensby:2014aa} whose accurate abundances have enabled a pioneering and comprehensive view of the chemical abundance structure in the Galactic disk. For completeness, we also note that in the differential analysis of solar twins by \citet{spina:2018aa}, the average differential errors are $\sigma$ \teff\ = 4.23 K, $\sigma$ \logg\ = 0.0114 dex and $\sigma$ [Fe/H] = 0.0037 dex. Their uncertainties are a factor of $\sim$3.5 better than in our study and the higher precision is driven, in part, by the very high quality spectra (R = 115,000, S/N $\ge$ 300 with a median value of 800) and the smaller difference in stellar parameters between stars in a given co-moving pair; average \teff\ and [Fe/H] differences for their sample are 132 K and 0.032 dex while for our sample the average differences in \teff\ and [Fe/H] within a given co-moving pair are 169 K and 0.056 dex. We then computed the abundances for all elements on a line-by-line differential basis and will present those results in Paper II (Liu et al.\ in prep). 

As noted, our analysis assumes 1D LTE. To assess any impact of 3D non-LTE, we use publicly available corrections from \citet{Amarsi:2022aa}. In our test, we selected a representative binary star pair with $\Delta$ \teff\ $\simeq$ 200K. Assuming that the stellar parameters were accurate (a strong assumption given that the they were derived using a line-by-line differential approach), we applied these corrections. The result was a minor average differential correction of 0.014 dex (with a standard deviation of 0.011 dex). This suggests that for pairs with temperature differences closer to 100K, the 3D non-LTE correction would be even smaller, potentially below 0.01 dex. It's crucial to emphasize that this test was based on one pair and the strong assumption of initial accuracy. However, it provides a preliminary insight into the magnitude of 3D non-LTE effects on our study's results. 

\section{Results} 

In addition to presenting our results in this section, we also provide a number of consistency checks to validate our stellar parameters and uncertainties. 

\subsection{Internal precision from multiple observations} 

For seven of the co-moving pairs, we have multiple (two) observations of both stars in the pair and we analysed each set of observations independently (as well as analysing the combined co-added observations). For five of the co-moving pairs, the multiple observations were taken with the Magellan Telescope on consecutive nights. For one of the co-moving pairs, one observation was taken with the Keck Telescope and the other observations with the Magellan Telescope. Finally, for the analysis of a co-moving pair, we ``swapped'' the reference star and object in the differential analysis. Therefore, these seven co-moving pairs enable us to check and quantify the internal precision of our results. 

\begin{figure}
    \centering
    \includegraphics[width=.75\hsize,angle=90]{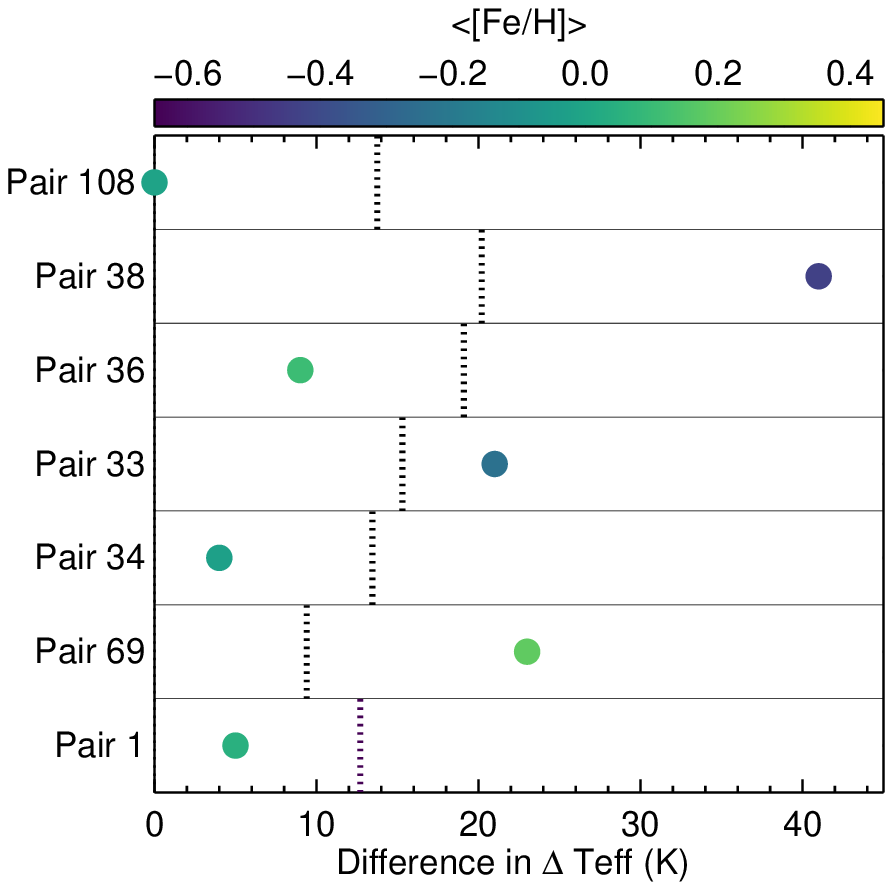}
    \caption{Differences in $\Delta$ \teff\ (reference $-$ object) from independent analysis of multiple observations of the same stars. The data are colour coded by the metallicity, [Fe/H]. The dashed vertical line indicates the average uncertainty.} 
    \label{fig:int_comp.1}
\end{figure}

\begin{figure}
    \centering
	\includegraphics[width=.75\hsize,angle=90]{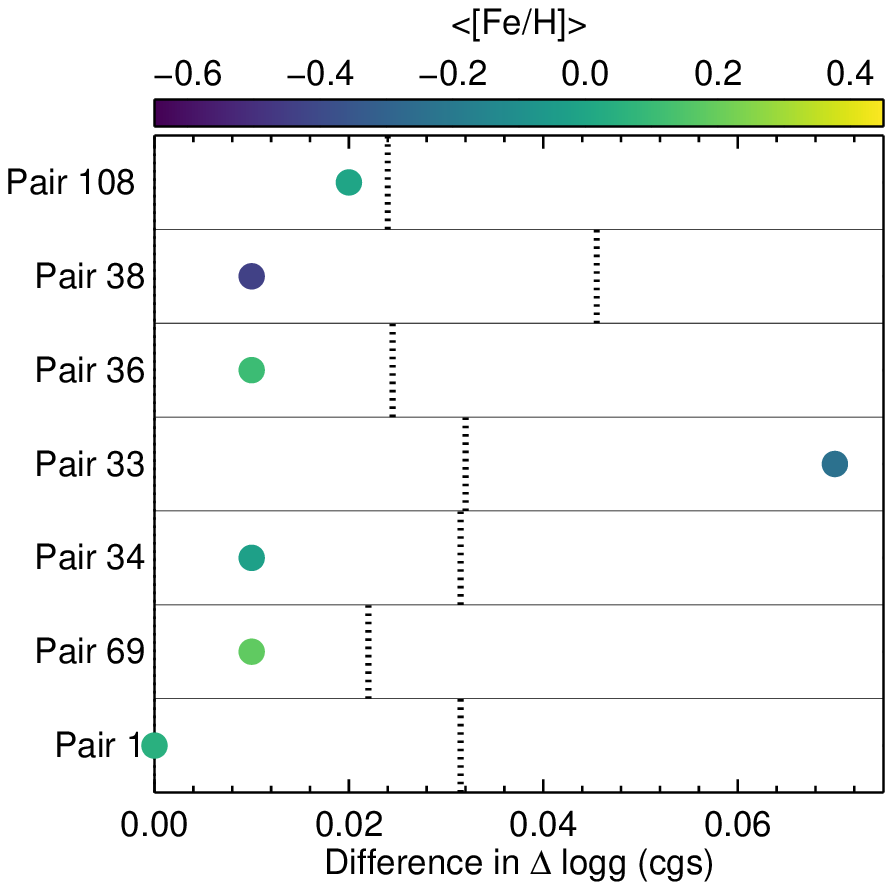}
    \caption{Same as Figure \ref{fig:int_comp.1} but for surface gravity, \logg.} 
    \label{fig:int_comp.2}
\end{figure}

\begin{figure}
    \centering
	\includegraphics[width=.75\hsize,angle=90]{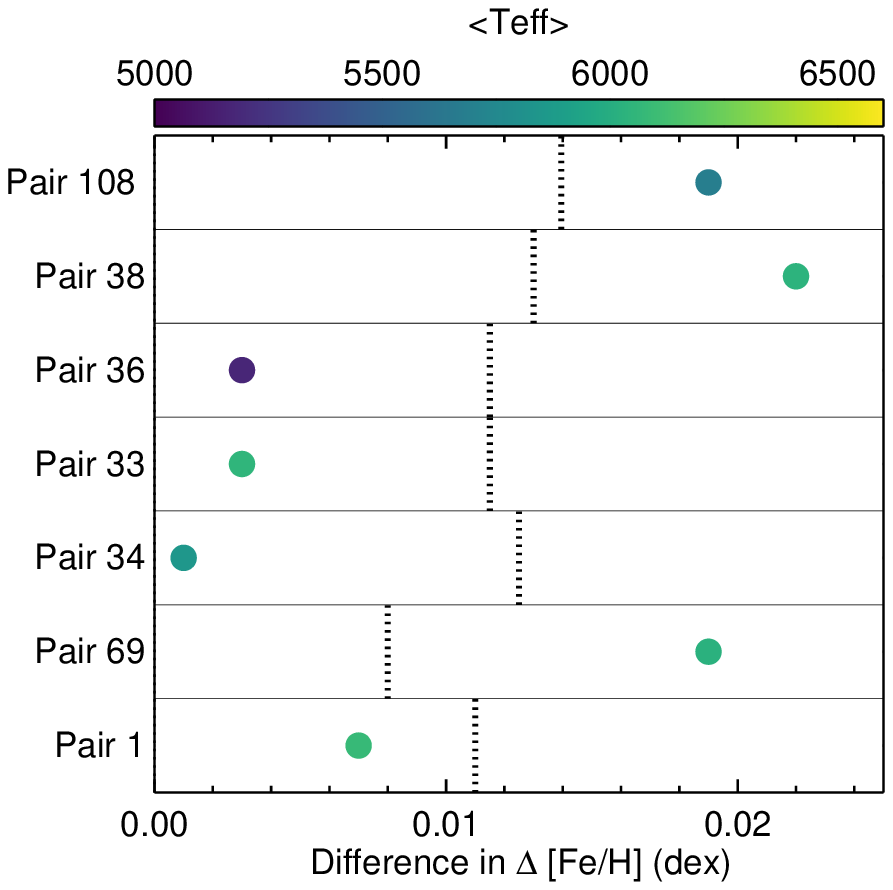}
    \caption{Same as Figure \ref{fig:int_comp.1} but for metallicity, [Fe/H], and colour coded by effective temperature, \teff.}
    \label{fig:int_comp.3}
\end{figure}

In Figures \ref{fig:int_comp.1}, \ref{fig:int_comp.2} and \ref{fig:int_comp.3} we present the differences in $\Delta$ \teff, $\Delta$ \logg\ and $\Delta$ [Fe/H], respectively, from the independent analyses of the multiple observations. The data are colour coded by [Fe/H] or \teff. Note that $\Delta$ refers to ``reference $-$ object'' for a given co-moving pair such that these figures are showing the `difference of differences' to quantify the internal precision of our differential analysis. For completeness, we note that the average absolute differences in $\Delta$ \teff, $\Delta$ \logg\ and $\Delta$ [Fe/H] for the seven co-moving pairs with multiple observations are 14.7 K, 0.019 dex and 0.011 dex, respectively, and these small values are in excellent agreement with the average errors for these quantities of 14.8 K, 0.030 dex and 0.012 dex. (Again, these values refer to the subset of stars for which multiple observations were analysed independently and not for the entire sample.) Therefore, this independent analysis of multiple observations of the co-moving pairs validates the fidelity of our stellar parameters and confirms the high precision nature of our differential analysis. Thus this comparison gives us confidence that our results are reliable and accurate. 

\subsection{Effective temperature} 

The effective temperature, \teff, of a star is one of the more readily measurable stellar parameters. However, direct measurements of \teff\ via angular diameter measurements are challenging and time consuming to obtain and calibrate at high precision, and are biased to the most nearby stars (e.g., see \citealt{huber:2012aa,rains:2020aa,tayar:2022aa}). While there are a few hundred stars with angular diameter measurements, only a fraction of these have been benchmarked more completely to have reliable ages, masses, and chemistry like the 34 \gaia\ FGK benchmark sample.

Indirect measurements of \teff\ can be obtained by using the infrared flux method \citep{blackwell:1977aa} and ``colour - temperature'' relations \citep{Alonso:1999aa,Ramirez:2005ab,Casagrande:2010aa}. Given that none of our program stars have direct \teff\ measurements, we therefore rely upon indirect \teff\ measurements to check and validate our values which were obtained using the differential spectroscopic approach. 

\begin{figure}
	\includegraphics[width=.75\hsize,angle=90]{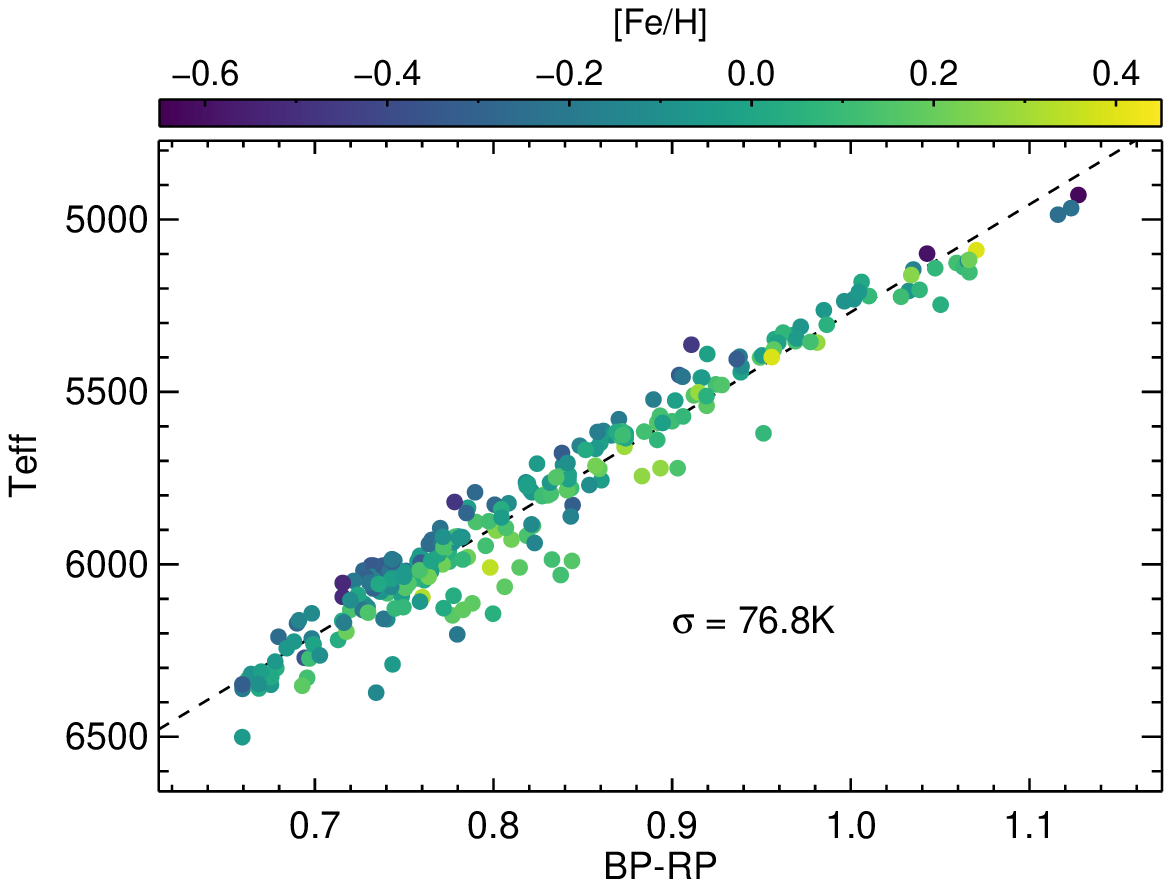}
    \caption{\teff\ vs.\ \bprp. The data are colour coded by metallicity and the dashed line represents the linear fit to the data. The dispersion about the linear fit is $\sigma$ = 76.8 K.}
    \label{fig:checks.1}
\end{figure}

In Figure \ref{fig:checks.1}, we present \teff\ versus the \bprp\ colour. Not surprisingly, there is a clear correlation between the two quantities. We refrain from overplotting the \citet{Casagrande:2021aa} colour - temperature relation since the reddening, metallicity and surface gravity will affect \teff. However, we note that for constant \logg\ (4.4), solar metallicity and zero reddening, their colour - temperature relation would pass through the majority our data. The main point to note from this figure is that we include a linear fit to the data and find that the dispersion about the best fit is 76.8 K which is slightly higher than the formal error of 54-66 K from \citet{Casagrande:2021aa}. (If we had adopted a quadratic function, the dispersion about the best fit would be essentially unchanged at 76.0 K). 

\begin{figure}
	\includegraphics[width=.75\hsize,angle=90]{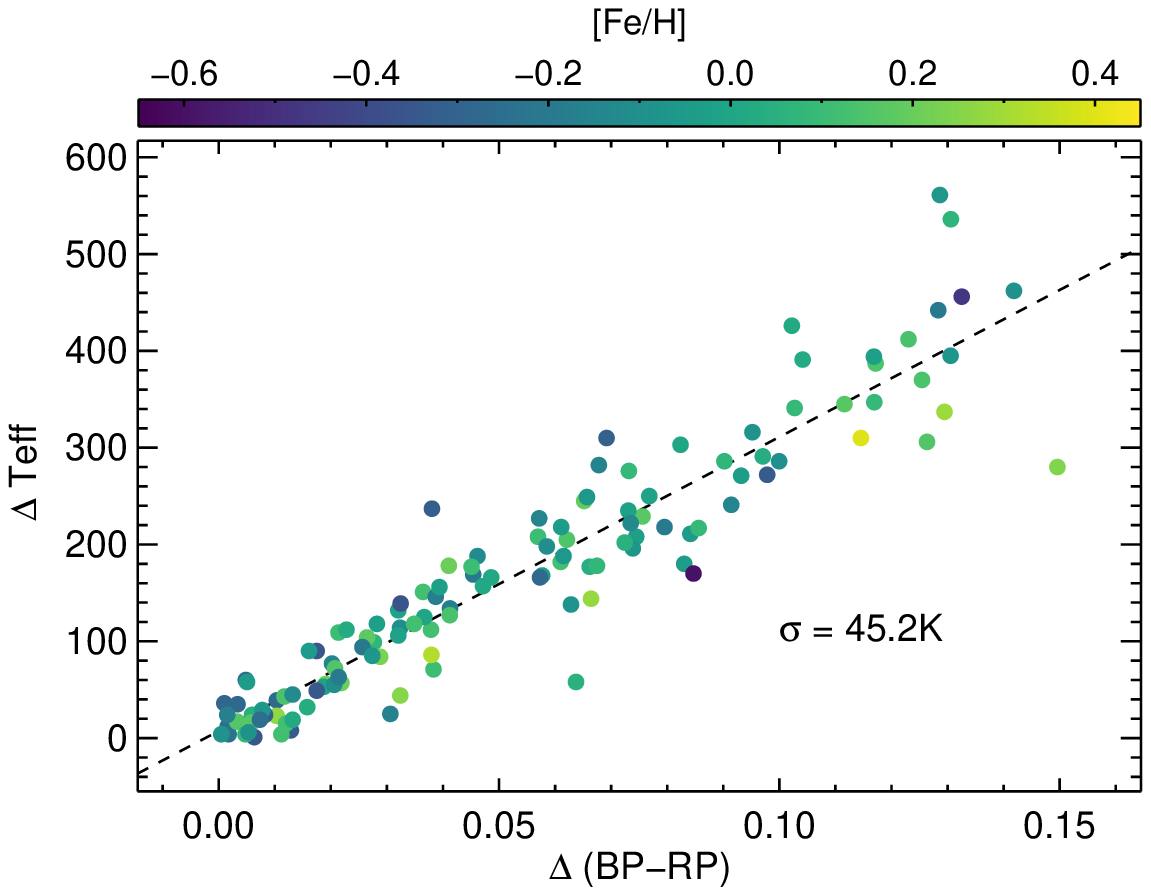}
    \caption{$\Delta$\teff\ vs.\ $\Delta$(\bprp). The data are colour coded by metallicity and the dashed line represents the linear fit to the data. The dispersion about the linear fit is $\sigma$ = 45.2 K.}
    \label{fig:checks.3}
\end{figure}

In Figure \ref{fig:checks.3}, we plot the difference in \teff\ versus the difference in \bprp\ for the two stars in a given co-moving pair. That is, this figure is the ``differential'' equivalent of Figure \ref{fig:checks.1}. The largest \teff\ differences between the stars in a given co-moving pair is nearly 600K. The most important aspect to note from this figure is that the dispersion about the linear fit to the data is only 45.2 K compared to 76.8 K in Figure \ref{fig:checks.1}. That is, when the difference in \teff\ is plotted against the difference in \bprp\, the dispersion about the linear fit is almost half relative to the absolute \teff:\bprp\ fit. Therefore, while Figure \ref{fig:checks.1} indicates that our results are accurate, Figure \ref{fig:checks.3} reveals that the precision increases considerably when we examine our stellar parameters from a differential perspective. 

\subsection{Evolutionary status} 

\begin{figure}
	\includegraphics[width=.75\hsize,angle=90]{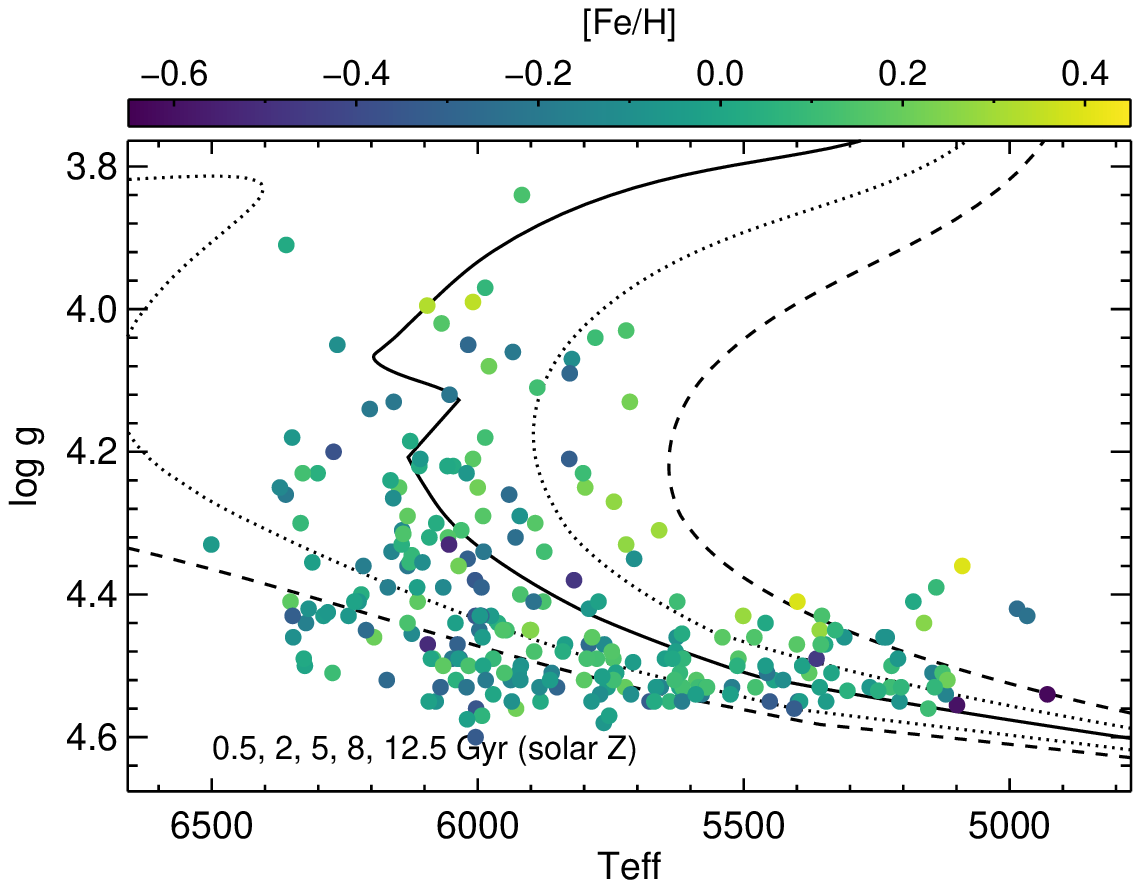}
    \caption{\teff\ vs.\ \logg. The data are colour coded by metallicity and the lines represents MIST isochrones of solar metallicity but for a range of ages from 0.5 Gyr (left-most line) to 12.5 Gyr (right-most line).}
    \label{fig:checks.4}
\end{figure}

In Figure \ref{fig:checks.4}, we plot our stars in the \teff\ versus \logg\ plane. In that figure we overplot MIST isochrones \citep{Choi:2016aa,dotter:2016aa} of solar metallicity with ages ranging from 0.5 Gyr to 12.5 Gyr. The program stars occupy plausible locations in this figure and as noted earlier, the majority are main sequence stars or subgiants, and ages could be obtained for some of these objects. 

\begin{figure}
	\includegraphics[width=.75\hsize,angle=90]{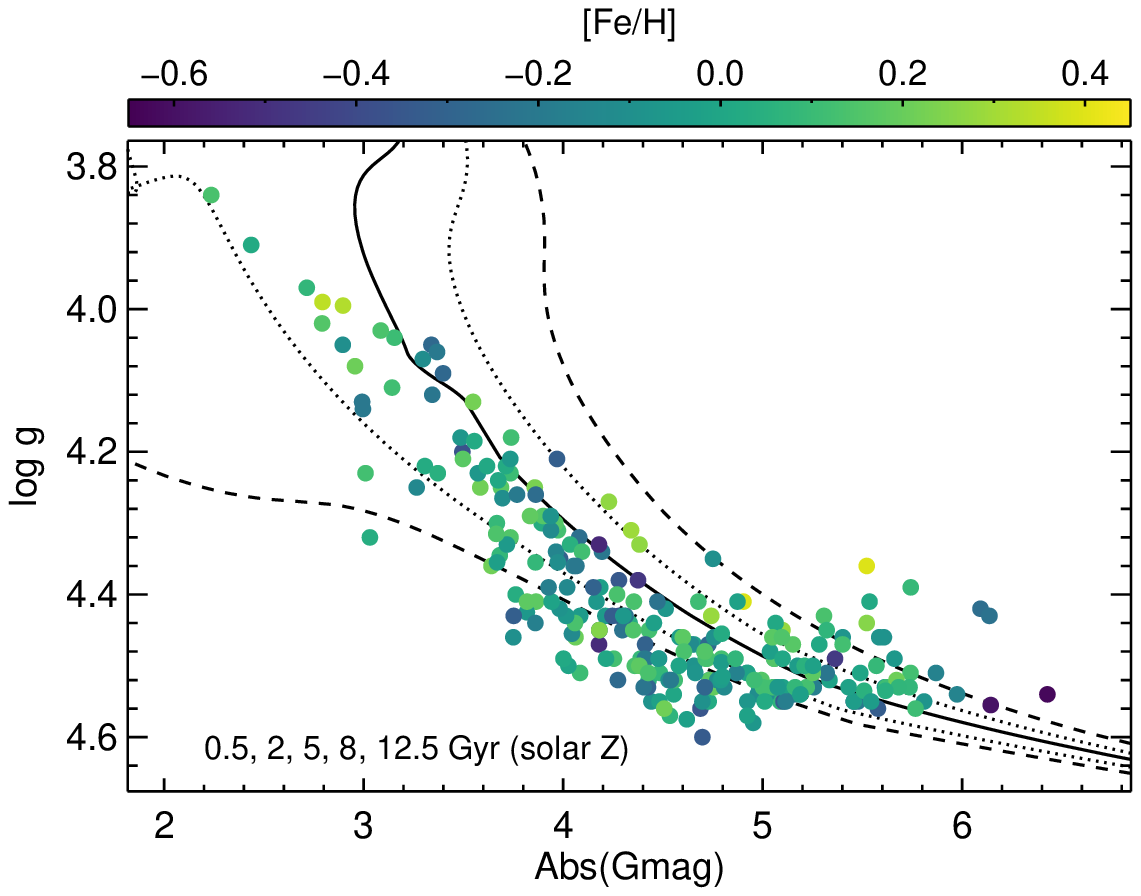}
    \caption{Absolute Gmag vs.\ \logg. The data are colour coded by metallicity and the lines represents MIST isochrones of solar metallicity but for a range of ages from 0.5 Gyr (lowest line) to 12.5 Gyr (highest line).}
    \label{fig:checks.6}
\end{figure}

In Figure \ref{fig:checks.6}, we plot our stars in the absolute Gmag\ versus \logg\ plane. As in Figure \ref{fig:checks.4} we overplot solar metallicity MIST isochrones with ages ranging from 0.5 Gyr to 12.5 Gyr and note that the program stars again lie on or near the isochrones. 

\subsection{Metallicity} 

\begin{figure}
	\includegraphics[width=.75\hsize,angle=90]{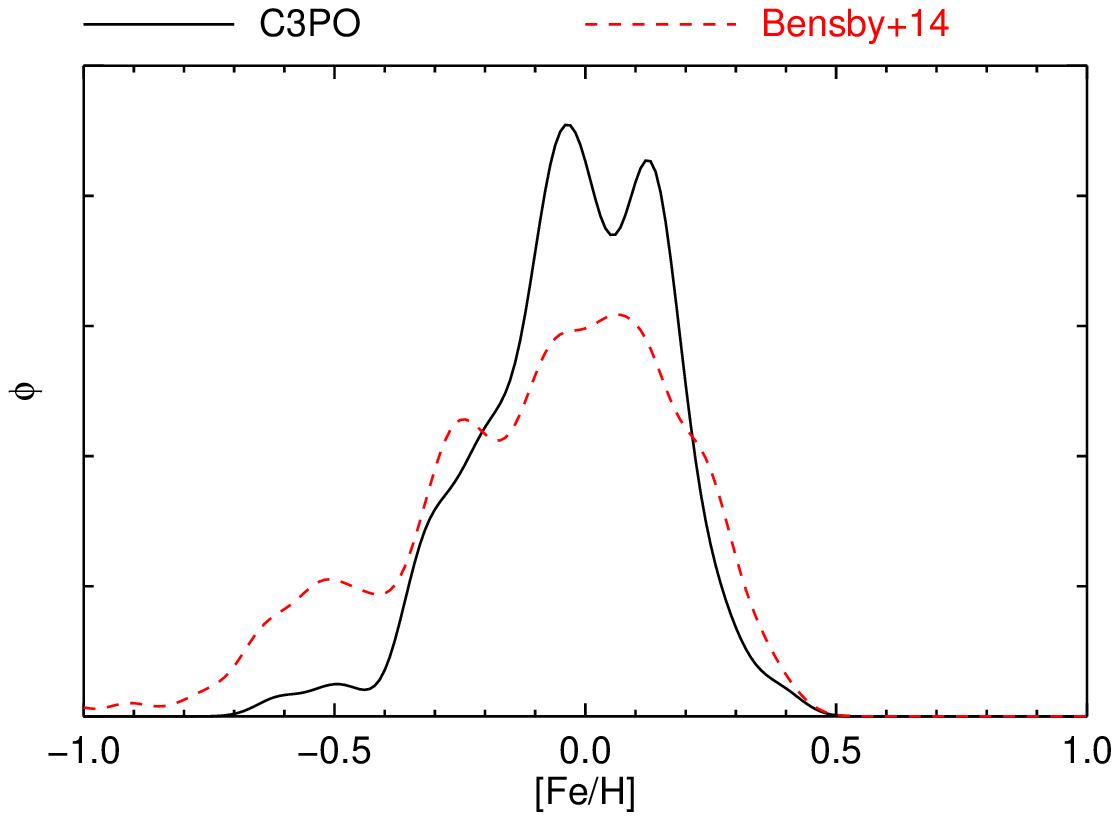}
    \caption{Metallicity distribution function for our sample (solid black line) and the \citet{bensby:2014aa} sample (red dashed line). }
    \label{fig:checks.26}
\end{figure}

Using iron as the canonical stellar measure of metallicity, we present the metallicity distribution function (MDF) in Figure \ref{fig:checks.26}. In that figure, we adopt  kernel density estimation, with a kernel size of 0.1 dex. As noted earlier, our data are centered near zero with tails to [Fe/H] = $\pm$0.5 dex. For comparison, we overplot the MDF from \citet{bensby:2014aa}; we chose their study as a comparison because their large sample (700 stars) are also primarily dwarfs and subgiants in the solar neighbourhood. (While their sample includes some 700 stars, we normalise both MDFs to have the same area.) The main difference between the two MDFs is that the \citet{bensby:2014aa} sample has more stars near [Fe/H] = $-$0.5, which presumably correspond to the thick disk. As they note in their paper, their study includes a very complex selection function designed to trace, among other things, ``the metal-poor limit of the thin disk, [and] the metal-rich limit of the thick disk.'' On the other hand, our sample is unlikely to have any metallicity bias (see Section 2 for details of the selection criteria) and is dominated by nearby objects, the majority of which presumably belong to the thin disk. 

\begin{figure*}
	\includegraphics[width=.5\hsize,angle=90]{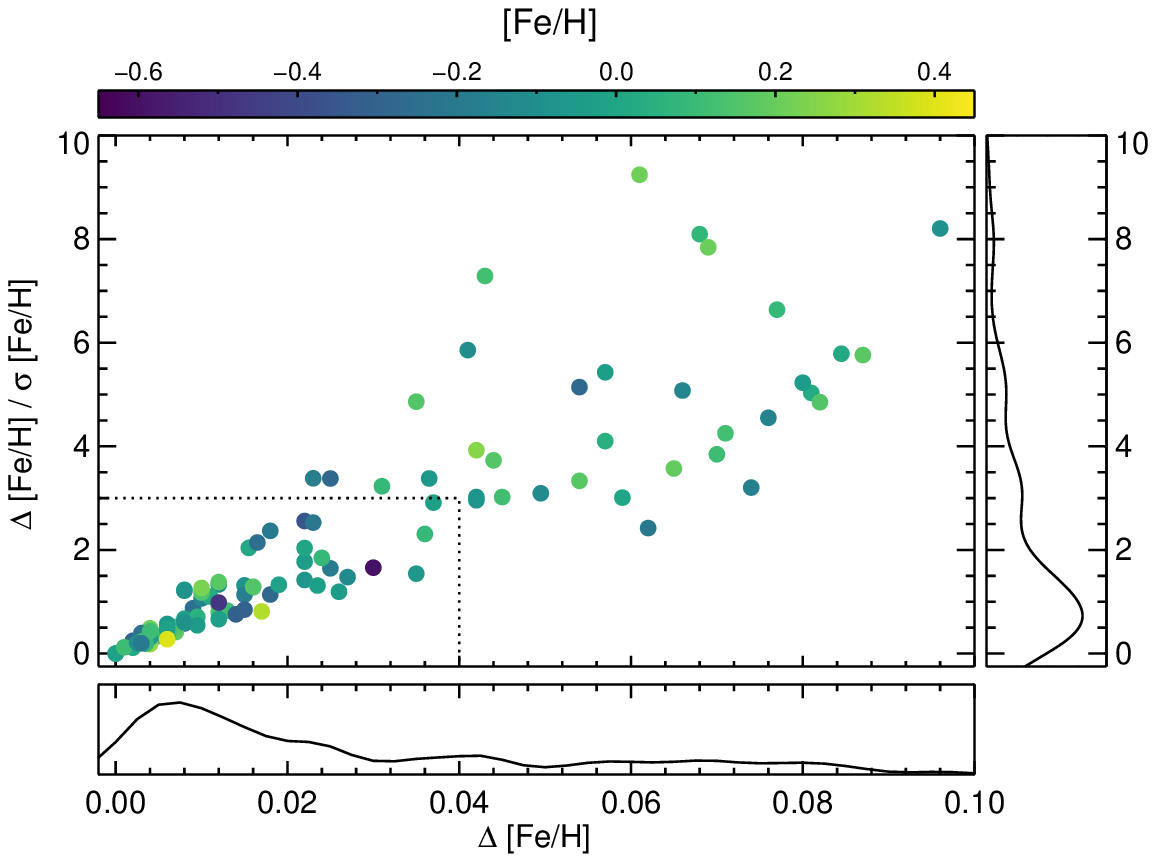}
    \caption{Difference in metallicity, $\Delta$[Fe/H] vs.\ the difference in metallicity normalised by the measurement uncertainty, $\Delta$[Fe/H] / $\sigma$[Fe/H]. The data are colour-coded by [Fe/H]. Marginal distributions are also included. The dashed lines indicate the sample we propose are ``chemically homogeneous'' (see text for details). } 
    \label{fig:checks.46}
\end{figure*}

One of the immediate goals of this study is to identify how to best split the sample into chemically homogeneous and inhomogeneous co-moving pairs to further explore the conclusions of \citet{Spina:2021aa}. In this context, we need to consider both the distribution in abundance differences in a given co-moving pair, $\Delta$[Fe/H], as well as the distribution of metallicity errors, $\sigma$[Fe/H]. In Figure \ref{fig:checks.46}, we plot the difference in [Fe/H] between the two stars in a given co-moving pair versus the same quantity but normalised by the uncertainty in [Fe/H]. In that figure, we focus on the 105 pairs with the $\Delta$[Fe/H] $\leq$ 0.1 dex rather than showing the full sample which extends up to $\Delta$[Fe/H] $\simeq$ 0.5 dex. 

Our assumption is that our sample of co-moving pairs can be separated into a subset that is chemically homogeneous with the remaining stars being regarded as chemically inhomogeneous. However, defining the sample of chemically homogeneous, i.e., ``normal'', co-moving pairs is non-trivial. For that subset of co-moving pairs, we expect the distribution of $\Delta$[Fe/H] / $\sigma$ [Fe/H] to be consistent with a Gaussian of width 1.0. Most of the dispersion in $\Delta$ [Fe/H] is driven by a few outliers. When culling the outliers with iterative sigma clipping we recover $\Delta$[Fe/H] $\simeq$ $\sigma$ [Fe/H], which further demonstrates that our uncertainties are robustly defined. We define the sample of chemically homogeneous co-moving pairs as those with $|\Delta$[Fe/H]$|$ / $\sigma$ [Fe/H] $\le$ 3.0 (i.e., a conservative approach) and $|\Delta$[Fe/H]$|$ $\le$ 0.04 dex (see Figure \ref{fig:checks.46}); there are 71 such pairs as well as 54 chemically inhomogeneous co-moving pairs. (We refrain from noting the percentage of chemically homogeneous pairs at this stage since we will impose additional constraints upon our sample in the following section.)

For completeness, we applied the same approach to the \citet{Spina:2021aa} data. Adopting their threshold of $|\Delta$[Fe/H]| / $\sigma$ [Fe/H] $\le$ 2.0, their  distribution of $\Delta$[Fe/H] / $\sigma$ [Fe/H] has a width of 1.0. Therefore, we confirm and validate their proposed separation of co-moving pairs but recognise that given the different data sets, the definition of chemically homogeneous co-moving pairs differs between the two samples.  

Finally, we note that \citet{Andrews:2019aa} examined a sample of wide binaries using APOGEE data and adopted $\Delta$ \teff\ $\le$ 200K as a definition of ``stellar twins''. If we follow their approach, 46 out of the 79 co-moving pairs are chemically inhomogeneous. 

\section{Discussion} 

Having defined our sample selection, described the analysis, conducted consistency checks and identified how to split the sample into homogeneous and inhomogeneous co-moving pairs, we now discuss our results in the context of the pilot study by \citet{Nelson:2021aa} and the analysis of over 100 binaries reported by \citet{Spina:2021aa}. The former paper investigated the chemical homogeneity as a function of spatial and kinematic separation among the co-moving pairs: our sample is a factor of four larger, 125 vs. 31 pairs of stars. The latter study reported evidence for an increase in the frequency of chemically inhomogeneous co-moving pairs with increasing average \teff. 

\subsection{Chemically homogeneity as a function of spatial and velocity separation.}

We now seek to build upon the results of the pilot study by \citet{Nelson:2021aa} who investigated the chemical homogeneity of co-moving pairs of stars as a function of their 3D pair separation (\ds) and $\Delta$ 3D velocity separation (\dv). Before moving on to our results, we highlight their findings. \citet{Nelson:2021aa} studied 31 co-moving pairs to understand whether such objects are conatal. They separated their sample into ``close co-moving pairs'' (spatial separations below 1 pc; 2 $\times$ $10^5$ AU: 17 pairs) and ``far co-moving pairs'' (spatial separations above 1pc: 14 pairs). Among the close co-moving pairs, the median abundance difference was $\Delta$[Fe/H] = 0.03 dex (standard deviation = 0.05 dex) and for the far co-moving pairs the median difference was $\Delta$[Fe/H] = 0.05 dex (standard deviation = 0.08 dex). Both values were substantially smaller than for random pairs, $\Delta$[Fe/H] = 0.16 dex (standard deviation = 0.23 dex), which were created by randomly ``assigning every star to another star, which is not its co-moving partner'' \citep{Nelson:2021aa}. The close co-moving pairs were believed to be conatal and exhibited small iron abundance differences. While the far co-moving pairs were more chemically heterogeneous than the close co-moving pairs, they were more homogeneous than random pairs. They suggested that the ``far co-moving pairs are a mixture of conatal pairs and chance alignments''. With our larger sample size (125 pairs vs.\ 31 pairs) and higher abundance precision (average Fe errors of 0.014 dex vs.\ 0.026 dex), we can re-examine chemical homogeneity as a function of spatial and velocity separation. 

\begin{figure*}
        \includegraphics[width=.95\hsize]{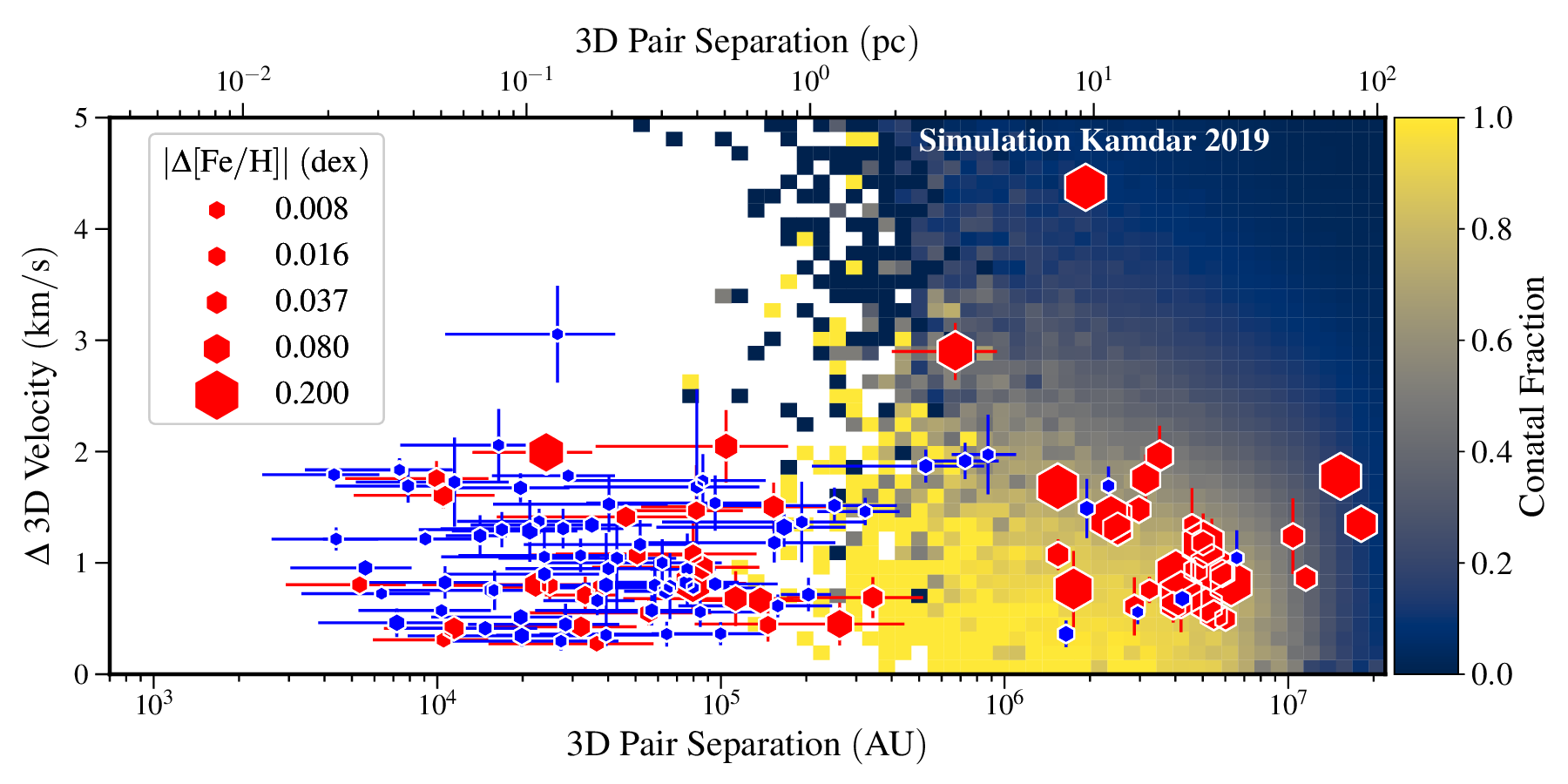}
    \caption{Velocity separation (\dv; \kms) vs.\ spatial separation (\ds; AU, pc) for the co-moving pairs of stars. The symbols sizes are scaled relative to the abundance difference between the two stars in a given pair. Chemically homogeneous and inhomogeneous co-moving pairs are denoted using blue and red symbols, respectively. Theoretical predictions for the conatal fraction from the \citet{Kamdar:2019aa} simulations are shown in the background.}
    \label{fig:anom_teff.boot2}
\end{figure*}

In Figure \ref{fig:anom_teff.boot2} we plot spatial separation \ds\ vs.\ velocity separation \dv\ for our sample. In that figure, the symbol size is proportional to the difference in metallicity between the two stars in a given co-moving pair, and chemically homogeneous and inhomogeneous pairs are coloured blue and red, respectively. When applying the spatial separation boundary of 1 pc (2 $\times$ $10^5$ AU) as used by \citep{Nelson:2021aa} and described above, we have 82 close co-moving pairs and 43 far co-moving pairs. The median of the absolute abundance difference for the close co-moving pairs is $\Delta$[Fe/H] = 0.016 dex (standard deviation = 0.035 dex) with an average uncertainty of $\sigma$[Fe/H] = 0.013 dex. For the far co-moving pairs, the median of the absolute abundance difference is $\Delta$[Fe/H] = 0.071 dex (standard deviation = 0.109 dex) with an average uncertainty of $\sigma$[Fe/H] = 0.016 dex. Following \citet{Nelson:2021aa}, we created a sample of random pairs and re-analysed these objects. The median of the absolute abundance difference is $\Delta$[Fe/H] = 0.038 dex (standard deviation = 0.215 dex) with an average uncertainty of 0.020 dex. Therefore, our results are qualitatively in agreement with \citet{Nelson:2021aa} in the sense that the standard deviation of the abundance differences is larger for the far co-moving pairs but smaller than for the random pairs. 

Inspection of the distribution of spatial separations in Figure \ref{fig:anom_teff.boot2} suggests a break around \ds\ = 10$^6$ AU ($\sim$ 4.8 pc). For spatial separations \ds\ $<$ 10$^6$ AU ($\sim$ 4.8 pc), the  majority of co-moving pairs are chemically homogeneous (63 out of 91 pairs; $\sim$70\%). For spatial separations higher than \ds\ = 10$^6$ AU, the minority of the co-moving pairs are chemically homogeneous (six out of 34 pairs; $\sim$18\%). Therefore, the first main conclusion is that we speculate that a spatial separation of \ds\ = 10$^6$ AU may represent a plausible boundary value to separate conatal and non-conatal pairs. For separations larger than \ds\ = 10$^6$ AU, we assume that chance alignments likely contaminate the sample although we agree with \citet{Kamdar:2019aa} that some of these objects are conatal. Therefore, moving forward in this paper, we will focus on the 91 co-moving pairs with \ds\ $\le$ 10$^6$ AU to more carefully investigate the properties of our sample. 

We report the following fractions of chemically inhomogeneous pairs for various ranges of \ds: (i) 20\% $\pm$ 15\% (two out of 10 for \ds\ $<$ 10$^4$ AU); (ii) 25\% $\pm$ 11\% (7 out of 28 for 10$^4$ $<$ \ds\ $\le$ 10$^{4.5}$ AU); (iii) 31\% $\pm$ 11\% (11 out of 35 for 10$^{4.5}$ $<$ \ds\ $\le$ 10$^5$ AU); (iv) 50\% $\pm$ 25\% (6 out of 12 for 10$^5$ $<$ \ds\ $\le$ 10$^{5.5}$ AU); and (v) 33\% $\pm$ 27\% (two out of six for 10$^{5.5}$ $<$ \ds\ $\le$ 10$^6$ AU). (We assume Poisson statistics in generating the uncertainties.) For a given co-moving pair, if the current separation of the two objects has remained the same throughout their evolution, then the fact that the chemically homogeneous fraction remains largely constant over three magnitudes of \ds\ would suggest that within a star forming region the ISM is homogeneous at scales up to 10 pc.

We also examine the fraction of chemically homogeneous co-moving pairs as a function of velocity separation, \dv. For co-moving pairs with \dv\ $\le$ 0.5 \kms, six of the 14 pairs are chemically inhomogeneous, i.e., 43\% $\pm$ 29\%. Similarly, we report the following values: 11 of the 33 pairs (33\% $\pm$ 16\%) are chemically inhomogeneous in the range 0.5 $<$ \dv\ $\le$ 1.0 \kms; four of the 21 pairs (19\% $\pm$ 14\%) are chemically inhomogeneous in the range 1.0 $<$ \dv\ $\le$ 1.5 \kms; and seven of the 23 pairs (30\% $\pm$ 13\%) are chemically inhomogeneous in the range \dv\ $\ge$ 1.5 \kms. (As above we assume Poisson statistics in generating the uncertainties.) Thus, our second main conclusion is that there appears to be no obvious trend of the fraction of chemically homogeneous co-moving pairs as a function of velocity separation, at least for the subset of objects with \ds\ $\le$ 10$^6$ AU and with \dv\ $\le$ 4 \kms. 

\subsection{Frequency of chemically inhomogeneous co-moving pairs as a function of effective temperature.}

\citet{Spina:2021aa} reported that the fraction of chemically inhomogeneous co-moving pairs of stars increased with increasing \teff\ and attributed this effect to planet ingestion. That is, stars with lower values of \teff\ have larger surface convection zones such that they can ingest a larger amount of rocky material from the planetary system, without a noticeable change in stellar atmospheric composition, relative to stars of higher \teff\ with smaller surface convection zones. In this scenario, stars with higher \teff\ which have ingested planetary material will exhibit larger abundances relative to (i) stars with cooler \teff\ and/or (ii) stars which have not ingested planets. For further details of planet ingestion, see N-body simulations by \citet{Izidoro:2021aa} and \citet{Bitsch:2023aa}.

\begin{figure*}
	\includegraphics[width=.5\hsize,angle=90]{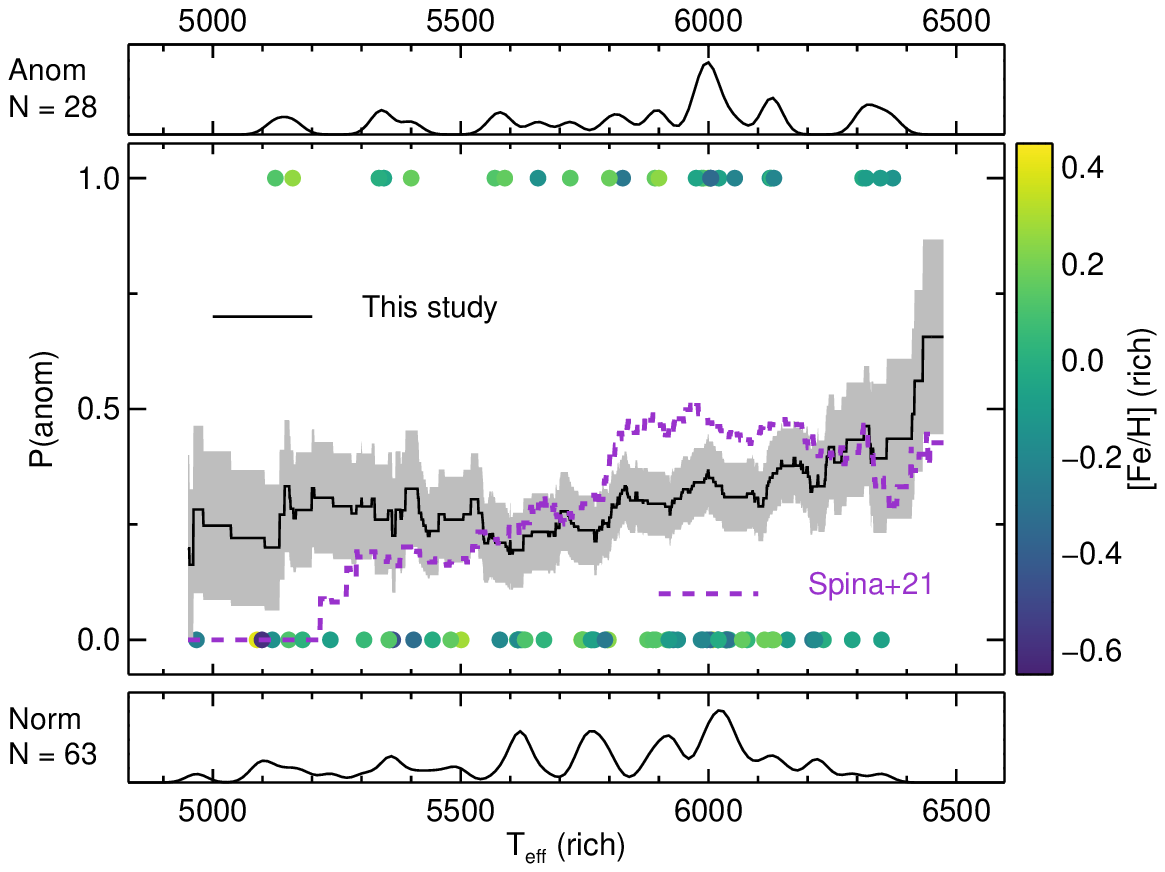}
    \caption{Frequency of chemically inhomogeneous co-moving pairs as a function of the \teff\ of the more metal-rich object. The \teff\ histograms for the chemically normal and inhomogeneous co-moving pairs are shown below and above, respectively. The data are colour coded by the  metallicity. The solid line and shaded grey regions represent the fraction of chemically inhomogeneous pairs within \teff\ $\pm$ 200K and the corresponding uncertainty, respectively. The \citet{Spina:2021aa} data are overplotted as the purple line.}
    \label{fig:anom_teff}
\end{figure*}

In Figure \ref{fig:anom_teff} we show the frequency of chemically inhomogeneous co-moving pairs as a function of \teff\ for the more metal-rich object in a given co-moving pair. In this figure, we restrict our sample to the co-moving pairs with spatial separation \ds\ $<$ 10$^6$ AU ($\sim$ 4.8 pc) for the reasons described above. The \teff\ distributions for the chemically normal and chemically inhomogeneous co-moving pairs are shown at the bottom and top of the figure, respectively. 

The solid black line in Figure \ref{fig:anom_teff} was produced by bootstrapping the data and applying a smoothing window of 200K. Thus for a given value of \teff, we consider the fraction of chemically inhomogeneous co-moving pairs within \teff\ $\pm$ 200K for all 10,000 iterations. The solid black line is the average of the distribution and the shaded region is the standard deviation. It is clear that the frequency of chemically inhomogeneous co-moving pairs increases with increasing \teff. For different values of \teff\ (reference star, object star or average), our conclusions are unchanged. Similarly, when considering co-moving pairs within \teff\ $\pm$ 50K or 100K (i.e., changing the smoothing window from 200K to 50K or 100K), our conclusions are unchanged. 

We then apply the same methodology to the \citet{Spina:2021aa} sample along with their definition for chemically inhomogeneous pairs. In Figure \ref{fig:anom_teff} we overplot their frequency of chemically inhomogeneous co-moving pairs as a purple line. While their sample exhibits an increase in the fraction of chemically inhomogeneous pairs with increasing \teff, their data exhibits the largest increase below \teff\ $\simeq$ 5750K then flattens out at higher \teff. Conversely, in our data the fraction of chemically inhomogeneous pairs appears to exhibit a flat trend below \teff\ $\simeq$ 5750K and then an increase at higher \teff. (If we applied a 2-$\sigma$ cut as per Spina et al.~our results would be qualitatively unchanged.) Given that \citet{Spina:2021aa} reported that the increase in the fraction of chemically inhomogeneous pairs is due to planet ingestion, which of these trends (if either) would be predicted? 

\begin{figure}
	\includegraphics[width=1.4\hsize,angle=90]{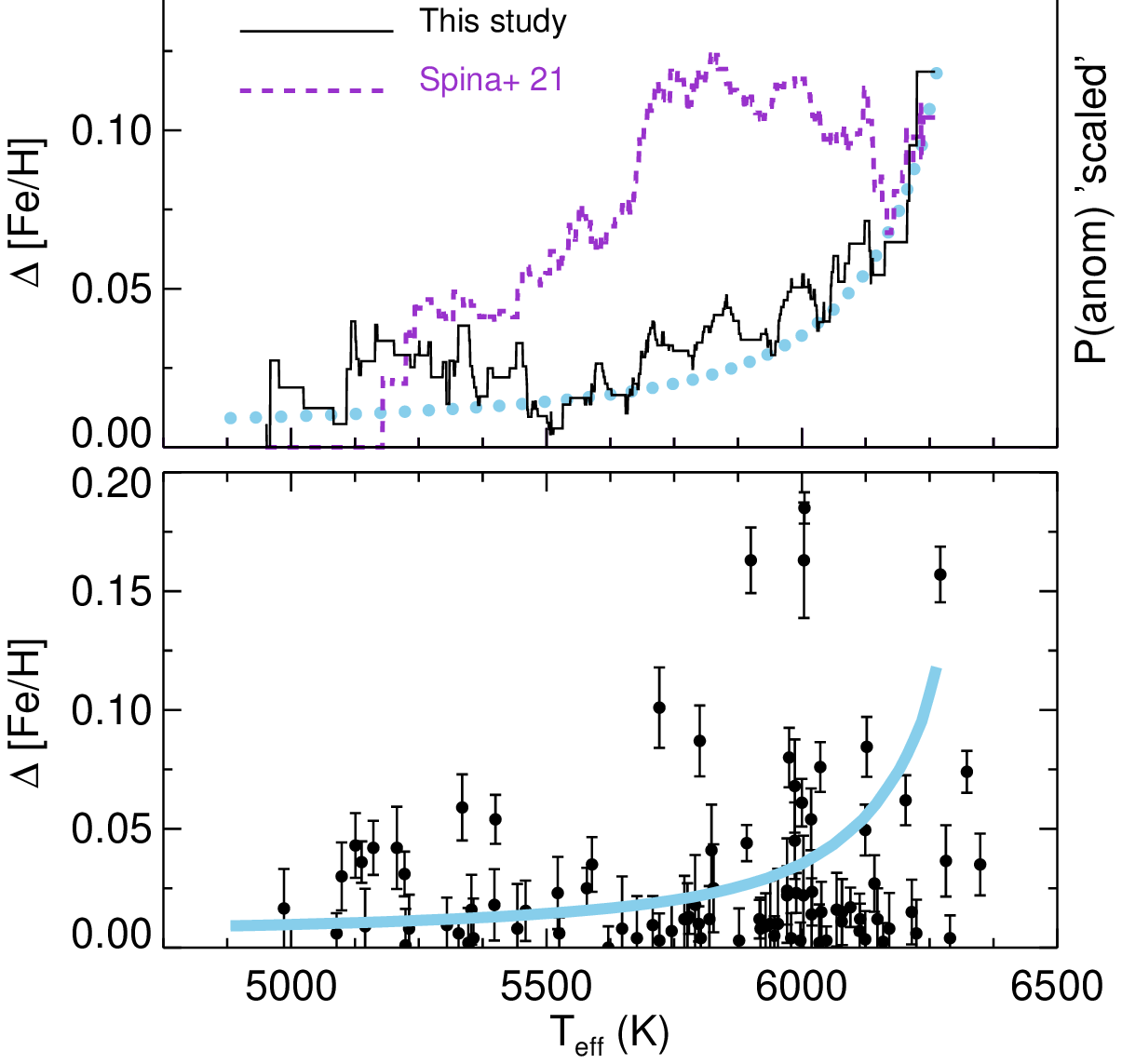}
    \caption{The upper panel shows mass of the convective envelope (\msun) versus \teff\ at the solar age (4.6 Gyr) and metallicity (Z=0.014; \citealt{Asplund09} abundance scale). The middle panel shows the change in [Fe/H] when 1 M$_{\earth}$ of material is injected into the convective envelope. The frequency of chemically inhomogeneous co-moving pairs of stars for this study (black) and \citeauthor{Spina:2021aa} (purple) are overplotted in the lower panel; both have been anchored and linearly stretched (see text for details). The lower panel shows the $\Delta$[Fe/H] values.
    \label{fig:mconv}}
\end{figure}

In the upper panel of Figure \ref{fig:mconv}, we plot the mass of the surface convection zone (i.e., convective envelope) as a function of \teff\ at an age of 4.6 Gyr and solar metallicity based on a grid of stellar evolution calculations performed with the Modules for Experiments in Stellar Astrophysics (MESA; \citealt{MESAI, MESAII, MESAIII, MESAIV, MESAV, MESAVI}) program version 21.12.1. The model grid assumes for all tracks a metal fraction of $Z=0.014$, a convective mixing length of $\alpha_\text{MLT} = 1.95$, the \citet{Asplund09} solar abundances, and the \texttt{photosphere} atmospheric boundary conditions, based on the MARCS model atmospheres \citep{Marcs}. The grid samples masses from 0.8 to 1.3 $M_{\odot}$ in increments of 0.01, and only those stars which have a convective envelope and radiative core at an age of 4.6 Gyr are considered in the figure (i.e., no models exhibiting core convection at this point are considered).

As \teff\ increases, the mass of the surface convection zone decreases as expected. Inspection of Figure 1 in \citet{Pinsonneault:2001aa} suggests that our values are in excellent agreement with theirs. While the mass of the surface convection zone decreases with increasing \teff, how does this affect the fraction of chemically inhomogeneous pairs? 
To explore this question, we performed the following test using a toy model. For each value of the mass of the surface convection zone, we assume that 1 M$_{\earth}$ of material is injected into the convective envelope (assuming solar metallicity) and calculated the change in [Fe/H]. That change in metallicity, $\Delta$[Fe/H] is plotted in the lower panel of Figure \ref{fig:mconv}. Note that the rate at which the metallicity increases with increasing \teff\ rises sharply around the solar value (\teff\ = 5772K). 

In the lower panel of Figure \ref{fig:mconv}, we overplot the fraction of chemically inhomogeneous stars from this study and from \citet{Spina:2021aa}. We anchor and linearly stretch those fractions\footnote{For our data: x' = 0.86x + 4950; y' = 1.3(y-0.17). For the Spina data: x' = 0.86x + 4950; y' = 1.3y.} to more closely match the predicted change in [Fe/H]. Clearly the change in metallicity as a function of \teff\ more closely resembles the increase in the fraction of inhomogeneous co-moving pairs shown by our data rather than that of \citet{Spina:2021aa}. While this similarity would suggest that the ingestion of planetary material could account for the increase in the fraction of chemically inhomogeneous pairs, further investigation of the ages, timescale for accretion, metallicity, amount of accreted material, etc., are needed. In a future paper (Liu et al.\ in prep.), we present a detailed investigation of potential chemical signatures of planet engulfment in our sample by examining the pattern of abundance differences in detail \citep{Melendez:2009ab} and fit those data with a planet engulfment model. 

Some key ideas such as the evolution of surface abundance changes as a function of time are explored in \citet{Behmard:2023aa}. In particular, \citet{Behmard:2023aa} noted that ``engulfment signatures are largest and longest lived for 1.1-1.2\msun\ stars, but no longer observable after $\sim$2 Gyr post-engulfment'' due primarily to thermohaline mixing \citep{Ulrich:1972aa}. That said, our sample includes stars with a range of masses for which the timescales for the engulfment events are unknown. 

Another aspect is that while \citet{Spina:2021aa} considered 107 co-moving pairs, their study combined their analysis of 31 pairs with literature values for 76 pairs. Of their literature sample, the following studies included at least 10 pairs; \citet{Desidera:2004aa,Desidera:2006aa}, \citet{Hawkins:2020aa} and \citet{nagar:2020aa}. One possibility is that the increase in the fraction of chemically inhomogeneous pairs could be driven by one (or more) of the subsamples, However, when we apply our analysis to each of the subsamples, we find that there is no individual study which is singularly responsible for the increase in the fraction of chemically inhomogeneous pairs with increasing \teff. 

Therefore, the third main conclusion from our study is that both our sample and the \citet{Spina:2021aa} sample exhibit an increase in the fraction of chemically inhomogeneous pairs with increasing \teff. However, for our homogeneous sample the shape of the increase in the fraction of chemically inhomogeneous pairs with increasing \teff\ is tantalisingly similar to what is predicted based on a toy model in which we inject material into the convective envelope; the fraction increases most sharply above solar \teff. We speculate that this results was only possible due to (i) our unbiased sample selection, (ii) large sample size and (iii) the high-precision homogeneous analysis.

\subsection{Signatures of atomic diffusion?} 

Atomic diffusion is a generic term used to describe a variety of mixing processes in the atmospheres of stars that affect the apparent chemical composition \citep{Dotter:2015aa}, and should be most noticeable on the main sequence. If the two stars in a given co-moving pair have different evolutionary status, atomic diffusion could potentially induce abundance differences within a co-moving pair. However, identifying the evolutionary phases requires an understanding of the birth masses and attributing any abundance differences to atomic diffusion would then assume that the stars in a given co-moving pair have not interacted. As noted in the introduction, several studies have reported evidence for atomic diffusion \citep{Korn:2007aa,Nordlander:2012aa,Souto:2019aa,Liu:2019aa,Liu:2021aa}. 

\begin{figure*}
	\includegraphics[width=.5\hsize,angle=90]{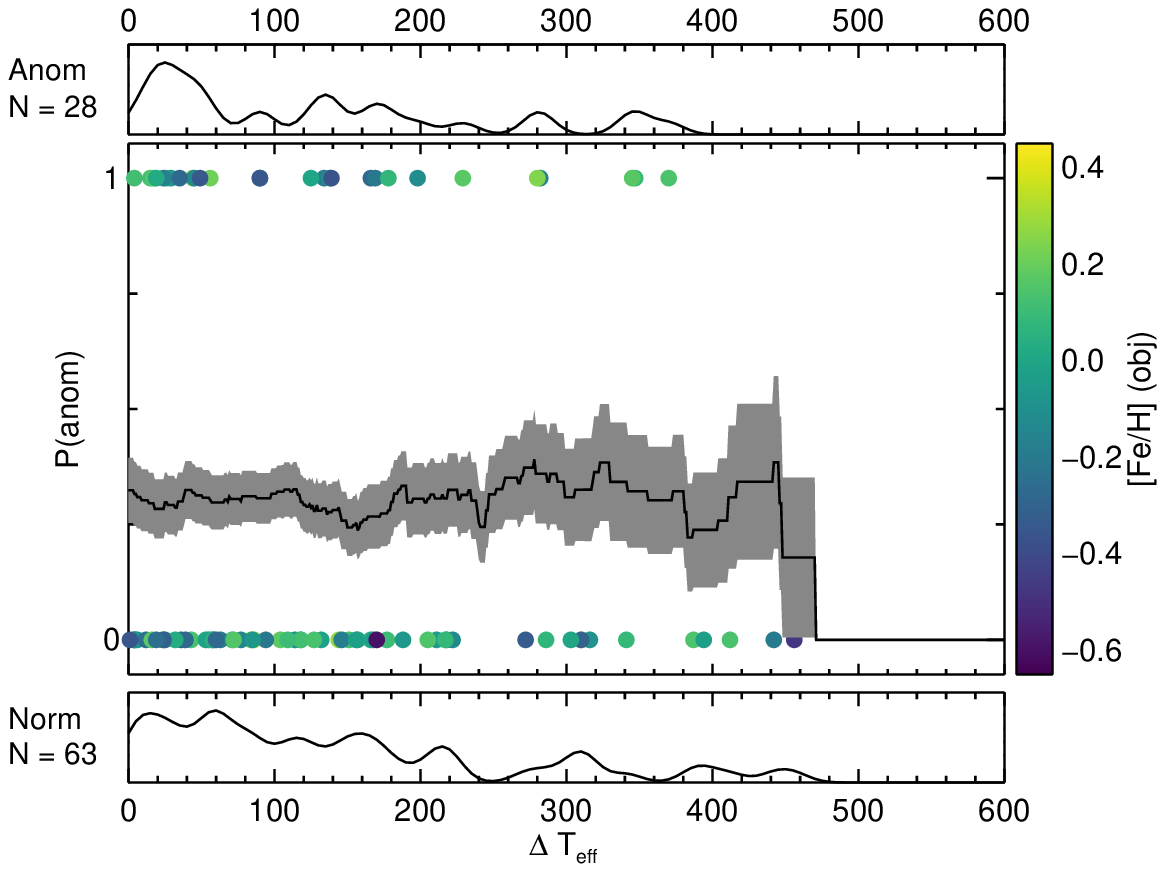}
    \caption{Frequency of chemically inhomogeneous co-moving pairs as a function of the \teff\ difference. The $\Delta$ \teff\ histograms for the chemically normal and inhomogeneous co-moving pairs are shown below and above, respectively. The data are colour coded by the  metallicity. The solid line and shaded grey regions represent the fraction of chemically inhomogeneous pairs within a smoothing window of \teff\ $\pm$ 100K and the corresponding uncertainty, respectively.}
    \label{fig:anom_delta_teff}
\end{figure*}

\begin{figure*}
	\includegraphics[width=.5\hsize,angle=90]{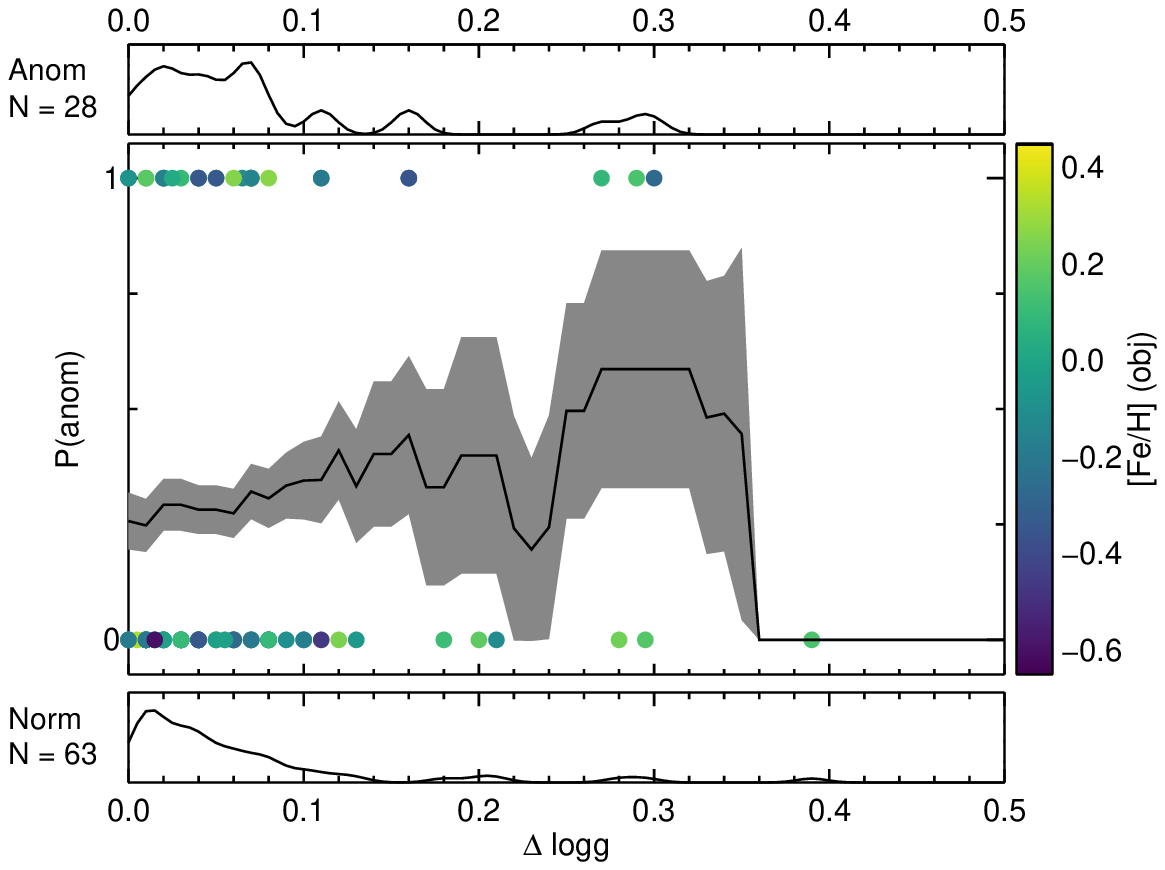}
    \caption{Same as Figure \ref{fig:anom_delta_teff} but for surface gravity, \logg. The solid line and shaded grey regions represent the fraction of chemically inhomogeneous pairs within a smoothing window of \logg\ $\pm$ 0.05 and the corresponding uncertainty, respectively.}
    \label{fig:anom_delta_logg}
\end{figure*}

In Figures \ref{fig:anom_delta_teff} and \ref{fig:anom_delta_logg}, we plot the frequency of chemically inhomogeneous pairs as a function of the difference in \teff\ and \logg, respectively. If the majority of chemically inhomogeneous pairs are due to atomic diffusion induced abundance differences, then we would expect to find the following trend; more chemically inhomogeneous pairs would be present when the differences in \teff\ and \logg\ between the two stars in a given pair are largest. Given the lack of a trend in Figure \ref{fig:anom_delta_teff} and the likely absence of a trend ($\sim$2-$\sigma$) in Figure \ref{fig:anom_delta_logg}, the fourth main conclusion is that atomic diffusion is unlikely to be the primary explanation for the majority of the chemically inhomogeneous pairs in our sample (based solely on Fe). 

Using the data from \citet{Dotter:2017aa} which corresponds to solar age and metallicity, we can undertake additional checks to understand the extent to which atomic diffusion could affect our sample. Using the surface gravities (or more precisely, the difference in surface gravity between the two stars in a given co-moving pair), the maximum change in iron abundance due to atomic diffusion in our sample is expected to be 0.077 dex. Recall that among the 91 pairs with co-moving pairs with \ds\ $\le$ 10$^6$ AU, we identified some 63 that were chemically homogeneous with $\Delta$[Fe/H] $\lesssim$ 0.04 dex. Based on the data from \citet{Dotter:2017aa}, 81 out of 91 pairs would be expected to have atomic diffusion induced abundance differences $\lesssim$ 0.04 dex, which confirms our earlier statement that atomic diffusion is unlikely to be the main explanation for the majority of chemically inhomogeneous pairs. That said, in a future paper in this series we will use the full suite of abundance measurements to further examine any role of atomic diffusion in our sample of co-moving pairs. 

\subsection{Implications and other considerations}

As noted in the introduction, co-moving pairs of stars may be used to calibrate and validate results from larger spectroscopic surveys. In this context, the 34 \gaia\ FGK ``benchmark'' stars \citep{Jofre:2014aa,Jofre:2015aa,Jofre:2017aa,Heiter:2015aa} and star clusters have played a significant role. However, we note that there is evidence that open clusters are not chemically homogeneous when high precision chemical abundance analyses are conducted \citep{Liu:2016ab,Liu:2016aa}. A key result from those studies is that the Hyades open cluster is chemically inhomogeneous at the $\sim$0.02 dex level (i.e., $\sim$ 5\%). Or said differently, the chemical homogeneity of the Hyades has a `floor' of 0.02 dex. 

In this study, we emphasise that some 56 co-moving pairs of stars exhibit abundance differences {\it below} the `Hyades floor', $\Delta$ [Fe/H] $\le$ 0.02 dex (i.e., $\le$ 5\%) of which some 34 co-moving pairs have abundance differences $\Delta$ [Fe/H] $\le$ 0.01 dex (i.e., $\le$ 2\%). Therefore, an immediate outcome of this study is a sample of bright nearby stars which are chemically identical at the $\le$ 2\% level which may be used to validate and calibrate chemical abundance results from large spectroscopic surveys. That is, each of these pairs is more chemically identical than the current members of the Hyades open cluster. 

On the other hand, if we assume that the 0.02 dex ``inhomogeneity floor'' of the Hyades means that chemical abundances are not homogeneous at the scale of the Hyades, our results would show that the chemistry is homogeneous at least at scales $<$ 10 pc. Extending the analysis to other elements might shed further light on this. 

In the lower panel of Figure \ref{fig:mconv}, the similarity between the change in [Fe/H] due to planetary material being injected into the convective envelope and the frequency of chemically inhomogeneous co-moving pairs of stars is striking. However, attributing this entirely to planet ingestion could imply that only one of the stars in a given co-moving pair has ingested a planet. If planet ingestion is a common process, and if that process does affect the stellar chemical composition, then the other star in a co-moving pair might also have ingested planets thereby erasing any abundance difference. The most likely planets to be ingested are super-Earths (or mini-Neptunes) since they orbit close to the star. Furthermore, these planets are in the mass range required to explain the chemical differences seen in Figure \ref{fig:mconv}. Indeed, while some 30-50\% of all stars likely host such planets \citep{Fressin:2013aa,Mulders:2018aa}, differences in the disc evolution scenario proposed by \citet{Hoppe:2020aa} might also play a role. In their scenario, each star has a disc which might have slightly different properties. The inward drifting solids that are then accreted onto the star could change the abundances of one of the objects in the co-moving pair. Another possibility is that the drifting solids could be blocked by growing planets in one of the objects \citep{Booth:2020aa}. In general, there should be outer (giant) planets that caused the inner ones to fall onto the host star, and such outer giant planets are rare \citep{Johnson:2010ab,Rosenthal:2022aa}. While there may be various explanations for the increase in the frequency of chemically inhomogeneous co-moving pairs of stars with increasing effective temperature, the causes we have noted all involve planets or planetary material.

\section{Conclusions}

This paper describes the sample selection, high-resolution spectroscopic observations, differential chemical abundance analysis of a sample of 125 co-moving pairs of stars. Recall that our definition of ``co-moving pairs of stars'' refers to both bound binary and unbound co-moving systems. These co-moving pairs include objects with large spatial separations which therefore offer an extension to classical binaries. 

To our knowledge, this work represents the largest high-precision chemical abundance study of such objects. The first step in this study was to confirm that a plausible boundary could be identified to distinguish between homogeneous and inhomogeneous co-moving pairs of stars; |$\Delta$[Fe/H]| $\le$ 0.04 dex. Our assumption is that the former are conatal while the latter are not. We then examined the fraction of chemically homogeneous co-moving pairs as a function of spatial separation, velocity separation and effective temperature. The four main conclusions from this study are: 

\begin{itemize}
    \item We speculate that a spatial separation of \ds\ = 10$^6$ AU may represent a boundary between homogeneous (i.e., conatal) and inhomogeneous (i.e., non-conatal) pairs of stars. For separations beyond \ds\ = 10$^6$ AU, we suggest that the sample are likely dominated by chance alignments. We restrict our conclusions to the 91 co-moving pairs with \ds\ $\le$ 10$^6$ AU and find that fraction of chemically homogeneous pairs is constant over three magnitudes of \ds. 
    \item There is no obvious trend between the fraction of chemically homogeneous co-moving pairs of stars and the velocity separation, at least within the range \ds\ $\le$ 10$^6$ AU and with \dv\ $\le$ 4 \kms\ which confirms and extends the work by \citet{Kamdar:2019aa}. 
    \item We verify an increase in the fraction of chemically inhomogeneous pairs with increasing \teff\ as reported by \citet{Spina:2021aa}. Our trend bears a strong similarity to the expected trend of planet ingestion from our toy model (but see caveats in Sec 5.4). That is, below solar \teff, the change in [Fe/H] increases mildly with increasing \teff\ while above solar \teff, [Fe/H] increases sharply with increasing \teff\ due to the decreasing mass of the convective envelope. 
    \item Atomic diffusion is unlikely to be the primary explanation for the majority of the chemically inhomogeneous pairs in our sample (although at this stage we are only considering the Fe abundance). 
\end{itemize} 

Another important outcome of this study is to provide the community with a sample of 56 bright co-moving pairs of stars which are chemically identical at the $\le$ 0.02 dex ($\sim$5\%) level, and these objects can be used to validate and calibrate abundance studies from larger surveys. This level of chemical homogeneity is comparable to that of the Hyades open cluster. The 34 co-moving pairs of stars which are chemically homogeneous at the $\le$ 0.01 dex ($\sim$ 2\%) level are important objects to facilitate calibrations from larger spectroscopic surveys. In the next papers in this series, we will consider the full set of element abundance ratios to identify evidence for planet engulfment among the co-moving pairs of stars.  

\section*{Acknowledgements}

The authors wish to recognize and acknowledge the very significant cultural role and reverence that the summit of Maunakea has always had within the indigenous Hawaiian community. We are most fortunate to have the opportunity to conduct observations from this mountain. We appreciate helpful comments from Adam Rains. We thank Tyler Nelson for the original version of Figure \ref{fig:anom_teff.boot2}. We appreciate helpful comments from the referee.

Parts of this research were supported by the Australian Research Council Centre of Excellence for All Sky Astrophysics in 3 Dimensions (ASTRO 3D), through project number CE170100013. Y.S.T.\ acknowledges financial support from the Australian Research Council through DECRA Fellowship DE220101520. M.J.\ gratefully acknowledges funding of MATISSE: \textit{Measuring Ages Through Isochrones, Seismology, and Stellar Evolution}, awarded through the European Commission's Widening Fellowship. This project has received funding from the European Union's Horizon 2020 research and innovation programme. B.B.~thanks the European Research Council (ERC Starting Grant 757448-PAMDORA) for their financial support. M.T.M.~acknowledges the support of the Australian Research Council through Future Fellowship grant FT180100194. 

This work has made use of data from the European Space Agency (ESA) mission
{\it Gaia} (\url{https://www.cosmos.esa.int/gaia}), processed by the {\it Gaia}
Data Processing and Analysis Consortium (DPAC,
\url{https://www.cosmos.esa.int/web/gaia/dpac/consortium}). Funding for the DPAC
has been provided by national institutions, in particular the institutions
participating in the {\it Gaia} Multilateral Agreement.

\section*{Data Availability}

The spectral data underlying this article are available in Keck Observatory Archive (\url{https://koa.ipac.caltech.edu/cgi-bin/KOA/nph-KOAlogin}) and ESO Science Archive Facility (\url{http://archive.eso.org/eso/eso_archive_main.html}). They can be accessed with Keck Program ID: W244Hr (Semester: 2021B, PI: Liu) and ESO Programme ID: 108.22EC.001, respectively. The data underlying this article will be shared on reasonable request to the corresponding author.



\bibliographystyle{mnras}

\begin{thebibliography}{}
\makeatletter
\relax
\def\mn@urlcharsother{\let\do\@makeother \do\$\do\&\do\#\do\^\do\_\do\%\do\~}
\def\mn@doi{\begingroup\mn@urlcharsother \@ifnextchar [ {\mn@doi@}
  {\mn@doi@[]}}
\def\mn@doi@[#1]#2{\def\@tempa{#1}\ifx\@tempa\@empty \href
  {http://dx.doi.org/#2} {doi:#2}\else \href {http://dx.doi.org/#2} {#1}\fi
  \endgroup}
\def\mn@eprint#1#2{\mn@eprint@#1:#2::\@nil}
\def\mn@eprint@arXiv#1{\href {http://arxiv.org/abs/#1} {{\tt arXiv:#1}}}
\def\mn@eprint@dblp#1{\href {http://dblp.uni-trier.de/rec/bibtex/#1.xml}
  {dblp:#1}}
\def\mn@eprint@#1:#2:#3:#4\@nil{\def\@tempa {#1}\def\@tempb {#2}\def\@tempc
  {#3}\ifx \@tempc \@empty \let \@tempc \@tempb \let \@tempb \@tempa \fi \ifx
  \@tempb \@empty \def\@tempb {arXiv}\fi \@ifundefined
  {mn@eprint@\@tempb}{\@tempb:\@tempc}{\expandafter \expandafter \csname
  mn@eprint@\@tempb\endcsname \expandafter{\@tempc}}}

\bibitem[\protect\citeauthoryear{{Alonso}, {Arribas}  \&
  {Mart{\'{\i}}nez-Roger}}{{Alonso} et~al.}{1999}]{Alonso:1999aa}
{Alonso} A.,  {Arribas} S.,   {Mart{\'{\i}}nez-Roger} C.,  1999, \mn@doi
  [\aaps] {10.1051/aas:1999521}, \href
  {http://adsabs.harvard.edu/abs/1999A%26AS..140..261A} {140, 261}

\bibitem[\protect\citeauthoryear{{Amarsi}, {Liljegren}  \& {Nissen}}{{Amarsi}
  et~al.}{2022}]{Amarsi:2022aa}
{Amarsi} A.~M.,  {Liljegren} S.,   {Nissen} P.~E.,  2022, \mn@doi [\aap]
  {10.1051/0004-6361/202244542}, \href
  {https://ui.adsabs.harvard.edu/abs/2022A&A...668A..68A} {668, A68}

\bibitem[\protect\citeauthoryear{{Andrews}, {Anguiano}, {Chanam{\'e}},
  {Ag{\"u}eros}, {Lewis}, {Hayes}  \& {Majewski}}{{Andrews}
  et~al.}{2019}]{Andrews:2019aa}
{Andrews} J.~J.,  {Anguiano} B.,  {Chanam{\'e}} J.,  {Ag{\"u}eros} M.~A.,
  {Lewis} H.~M.,  {Hayes} C.~R.,   {Majewski} S.~R.,  2019, \mn@doi [\apj]
  {10.3847/1538-4357/aaf502}, \href
  {https://ui.adsabs.harvard.edu/abs/2019ApJ...871...42A} {871, 42}

\bibitem[\protect\citeauthoryear{{Asplund}, {Grevesse}, {Sauval}  \&
  {Scott}}{{Asplund} et~al.}{2009}]{Asplund09}
{Asplund} M.,  {Grevesse} N.,  {Sauval} A.~J.,   {Scott} P.,  2009, \mn@doi
  [\araa] {10.1146/annurev.astro.46.060407.145222}, \href
  {http://adsabs.harvard.edu/abs/2009ARA%26A..47..481A} {47, 481}

\bibitem[\protect\citeauthoryear{{Behmard}, {Dai}, {Brewer}, {Berger}  \&
  {Howard}}{{Behmard} et~al.}{2023}]{Behmard:2023aa}
{Behmard} A.,  {Dai} F.,  {Brewer} J.~M.,  {Berger} T.~A.,   {Howard} A.~W.,
  2023, \mn@doi [\mnras] {10.1093/mnras/stad745}, \href
  {https://ui.adsabs.harvard.edu/abs/2023MNRAS.521.2969B} {521, 2969}

\bibitem[\protect\citeauthoryear{{Bensby}, {Feltzing}  \& {Oey}}{{Bensby}
  et~al.}{2014}]{bensby:2014aa}
{Bensby} T.,  {Feltzing} S.,   {Oey} M.~S.,  2014, \mn@doi [\aap]
  {10.1051/0004-6361/201322631}, \href
  {http://adsabs.harvard.edu/abs/2014A%26A...562A..71B} {562, A71}

\bibitem[\protect\citeauthoryear{{Bernstein}, {Shectman}, {Gunnels},
  {Mochnacki}  \& {Athey}}{{Bernstein} et~al.}{2003}]{Bernstein:2003aa}
{Bernstein} R.,  {Shectman} S.~A.,  {Gunnels} S.~M.,  {Mochnacki} S.,   {Athey}
  A.~E.,  2003, in {Iye} M.,  {Moorwood} A.~F.~M.,  eds,  \procspie Vol. 4841,
  Instrument Design and Performance for Optical/Infrared Ground-based
  Telescopes. pp 1694--1704

\bibitem[\protect\citeauthoryear{{Bitsch} \& {Izidoro}}{{Bitsch} \&
  {Izidoro}}{2023}]{Bitsch:2023aa}
{Bitsch} B.,  {Izidoro} A.,  2023, \mn@doi [arXiv e-prints]
  {10.48550/arXiv.2304.12758}, \href
  {https://ui.adsabs.harvard.edu/abs/2023arXiv230412758B} {p. arXiv:2304.12758}

\bibitem[\protect\citeauthoryear{{Bitsch}, {Forsberg}, {Liu}  \&
  {Johansen}}{{Bitsch} et~al.}{2018}]{Bitsch:2018aa}
{Bitsch} B.,  {Forsberg} R.,  {Liu} F.,   {Johansen} A.,  2018, \mn@doi
  [\mnras] {10.1093/mnras/sty1710}, \href
  {https://ui.adsabs.harvard.edu/abs/2018MNRAS.479.3690B} {479, 3690}

\bibitem[\protect\citeauthoryear{{Blackwell} \& {Shallis}}{{Blackwell} \&
  {Shallis}}{1977}]{blackwell:1977aa}
{Blackwell} D.~E.,  {Shallis} M.~J.,  1977, \mn@doi [\mnras]
  {10.1093/mnras/180.2.177}, \href
  {https://ui.adsabs.harvard.edu/abs/1977MNRAS.180..177B} {180, 177}

\bibitem[\protect\citeauthoryear{{Booth} \& {Owen}}{{Booth} \&
  {Owen}}{2020}]{Booth:2020aa}
{Booth} R.~A.,  {Owen} J.~E.,  2020, \mn@doi [\mnras] {10.1093/mnras/staa578},
  \href {https://ui.adsabs.harvard.edu/abs/2020MNRAS.493.5079B} {493, 5079}

\bibitem[\protect\citeauthoryear{{Brauer}, {Dullemond}  \& {Henning}}{{Brauer}
  et~al.}{2008}]{Brauer:2008aa}
{Brauer} F.,  {Dullemond} C.~P.,   {Henning} T.,  2008, \mn@doi [\aap]
  {10.1051/0004-6361:20077759}, \href
  {https://ui.adsabs.harvard.edu/abs/2008A&A...480..859B} {480, 859}

\bibitem[\protect\citeauthoryear{{Casagrande}, {Ram{\'{\i}}rez},
  {Mel{\'e}ndez}, {Bessell}  \& {Asplund}}{{Casagrande}
  et~al.}{2010}]{Casagrande:2010aa}
{Casagrande} L.,  {Ram{\'{\i}}rez} I.,  {Mel{\'e}ndez} J.,  {Bessell} M.,
  {Asplund} M.,  2010, \mn@doi [\aap] {10.1051/0004-6361/200913204}, \href
  {http://adsabs.harvard.edu/abs/2010A%26A...512A..54C} {512, A54}

\bibitem[\protect\citeauthoryear{{Casagrande} et~al.,}{{Casagrande}
  et~al.}{2021}]{Casagrande:2021aa}
{Casagrande} L.,  et~al., 2021, \mn@doi [\mnras] {10.1093/mnras/stab2304},
  \href {https://ui.adsabs.harvard.edu/abs/2021MNRAS.507.2684C} {507, 2684}

\bibitem[\protect\citeauthoryear{{Castelli} \& {Kurucz}}{{Castelli} \&
  {Kurucz}}{2003}]{Castelli:2003aa}
{Castelli} F.,  {Kurucz} R.~L.,  2003, in {Piskunov} N.,  {Weiss} W.~W.,
  {Gray} D.~F.,  eds,  IAU Symposium Vol. 210, Modelling of Stellar
  Atmospheres. p.~20P

\bibitem[\protect\citeauthoryear{{Chambers}}{{Chambers}}{2010}]{Chambers:2010aa}
{Chambers} J.~E.,  2010, \mn@doi [\apj] {10.1088/0004-637X/724/1/92}, \href
  {https://ui.adsabs.harvard.edu/abs/2010ApJ...724...92C} {724, 92}

\bibitem[\protect\citeauthoryear{{Choi}, {Dotter}, {Conroy}, {Cantiello},
  {Paxton}  \& {Johnson}}{{Choi} et~al.}{2016}]{Choi:2016aa}
{Choi} J.,  {Dotter} A.,  {Conroy} C.,  {Cantiello} M.,  {Paxton} B.,
  {Johnson} B.~D.,  2016, \mn@doi [\apj] {10.3847/0004-637X/823/2/102}, \href
  {https://ui.adsabs.harvard.edu/abs/2016ApJ...823..102C} {823, 102}

\bibitem[\protect\citeauthoryear{{De Silva} et~al.,}{{De Silva}
  et~al.}{2015}]{DeSilva:2015aa}
{De Silva} G.~M.,  et~al., 2015, \mn@doi [\mnras] {10.1093/mnras/stv327}, \href
  {https://ui.adsabs.harvard.edu/abs/2015MNRAS.449.2604D} {449, 2604}

\bibitem[\protect\citeauthoryear{{Dekker}, {D'Odorico}, {Kaufer}, {Delabre}  \&
  {Kotzlowski}}{{Dekker} et~al.}{2000}]{Dekker:2000aa}
{Dekker} H.,  {D'Odorico} S.,  {Kaufer} A.,  {Delabre} B.,   {Kotzlowski} H.,
  2000, in {Iye} M.,  {Moorwood} A.~F.,  eds,  \procspie Vol. 4008, Optical and
  IR Telescope Instrumentation and Detectors. pp 534--545,
  \mn@doi{10.1117/12.395512}

\bibitem[\protect\citeauthoryear{{Desidera} et~al.,}{{Desidera}
  et~al.}{2004}]{Desidera:2004aa}
{Desidera} S.,  et~al., 2004, \mn@doi [\aap] {10.1051/0004-6361:20041242},
  \href {https://ui.adsabs.harvard.edu/abs/2004A&A...420..683D} {420, 683}

\bibitem[\protect\citeauthoryear{{Desidera}, {Gratton}, {Lucatello}  \&
  {Claudi}}{{Desidera} et~al.}{2006}]{Desidera:2006aa}
{Desidera} S.,  {Gratton} R.~G.,  {Lucatello} S.,   {Claudi} R.~U.,  2006,
  \mn@doi [\aap] {10.1051/0004-6361:20064896}, \href
  {https://ui.adsabs.harvard.edu/abs/2006A&A...454..581D} {454, 581}

\bibitem[\protect\citeauthoryear{{Dotter}}{{Dotter}}{2016}]{dotter:2016aa}
{Dotter} A.,  2016, \mn@doi [\apjs] {10.3847/0067-0049/222/1/8}, \href
  {https://ui.adsabs.harvard.edu/abs/2016ApJS..222....8D} {222, 8}

\bibitem[\protect\citeauthoryear{{Dotter}, {Ferguson}, {Conroy}, {Milone},
  {Marino}  \& {Yong}}{{Dotter} et~al.}{2015}]{Dotter:2015aa}
{Dotter} A.,  {Ferguson} J.~W.,  {Conroy} C.,  {Milone} A.~P.,  {Marino} A.~F.,
    {Yong} D.,  2015, \mn@doi [\mnras] {10.1093/mnras/stu2170}, \href
  {http://adsabs.harvard.edu/abs/2015MNRAS.446.1641D} {446, 1641}

\bibitem[\protect\citeauthoryear{{Dotter}, {Conroy}, {Cargile}  \&
  {Asplund}}{{Dotter} et~al.}{2017}]{Dotter:2017aa}
{Dotter} A.,  {Conroy} C.,  {Cargile} P.,   {Asplund} M.,  2017, \mn@doi [\apj]
  {10.3847/1538-4357/aa6d10}, \href
  {https://ui.adsabs.harvard.edu/abs/2017ApJ...840...99D} {840, 99}

\bibitem[\protect\citeauthoryear{{El-Badry}, {Rix}  \& {Heintz}}{{El-Badry}
  et~al.}{2021}]{elbadry:2021aa}
{El-Badry} K.,  {Rix} H.-W.,   {Heintz} T.~M.,  2021, \mn@doi [\mnras]
  {10.1093/mnras/stab323}, \href
  {https://ui.adsabs.harvard.edu/abs/2021MNRAS.506.2269E} {506, 2269}

\bibitem[\protect\citeauthoryear{{Epstein}, {Johnson}, {Dong}, {Udalski},
  {Gould}  \& {Becker}}{{Epstein} et~al.}{2010}]{epstein:2010aa}
{Epstein} C.~R.,  {Johnson} J.~A.,  {Dong} S.,  {Udalski} A.,  {Gould} A.,
  {Becker} G.,  2010, \mn@doi [\apj] {10.1088/0004-637X/709/1/447}, \href
  {https://ui.adsabs.harvard.edu/abs/2010ApJ...709..447E} {709, 447}

\bibitem[\protect\citeauthoryear{{Feng} \& {Krumholz}}{{Feng} \&
  {Krumholz}}{2014}]{Feng:2014aa}
{Feng} Y.,  {Krumholz} M.~R.,  2014, \mn@doi [\nat] {10.1038/nature13662},
  \href {https://ui.adsabs.harvard.edu/abs/2014Natur.513..523F} {513, 523}

\bibitem[\protect\citeauthoryear{{Fressin} et~al.,}{{Fressin}
  et~al.}{2013}]{Fressin:2013aa}
{Fressin} F.,  et~al., 2013, \mn@doi [\apj] {10.1088/0004-637X/766/2/81}, \href
  {https://ui.adsabs.harvard.edu/abs/2013ApJ...766...81F} {766, 81}

\bibitem[\protect\citeauthoryear{{Freudling}, {Romaniello}, {Bramich},
  {Ballester}, {Forchi}, {Garc{\'{\i}}a-Dabl{\'o}}, {Moehler}  \&
  {Neeser}}{{Freudling} et~al.}{2013}]{Freudling:2013aa}
{Freudling} W.,  {Romaniello} M.,  {Bramich} D.~M.,  {Ballester} P.,  {Forchi}
  V.,  {Garc{\'{\i}}a-Dabl{\'o}} C.~E.,  {Moehler} S.,   {Neeser} M.~J.,  2013,
  \mn@doi [\aap] {10.1051/0004-6361/201322494}, \href
  {http://adsabs.harvard.edu/abs/2013A%26A...559A..96F} {559, A96}

\bibitem[\protect\citeauthoryear{{Gaia Collaboration} et~al.,}{{Gaia
  Collaboration} et~al.}{2016}]{gaia}
{Gaia Collaboration} et~al., 2016, \mn@doi [\aap]
  {10.1051/0004-6361/201629272}, \href
  {https://ui.adsabs.harvard.edu/abs/2016A&A...595A...1G} {595, A1}

\bibitem[\protect\citeauthoryear{{Gaia Collaboration} et~al.,}{{Gaia
  Collaboration} et~al.}{2021}]{gaiaedr3}
{Gaia Collaboration} et~al., 2021, \mn@doi [\aap]
  {10.1051/0004-6361/202039657}, \href
  {https://ui.adsabs.harvard.edu/abs/2021A&A...649A...1G} {649, A1}

\bibitem[\protect\citeauthoryear{{Gilmore} et~al.,}{{Gilmore}
  et~al.}{2012}]{Gilmore:2012aa}
{Gilmore} G.,  et~al., 2012, The Messenger, \href
  {https://ui.adsabs.harvard.edu/abs/2012Msngr.147...25G} {147, 25}

\bibitem[\protect\citeauthoryear{{Gustafsson}, {Edvardsson}, {Eriksson},
  {J{\o}rgensen}, {Nordlund}  \& {Plez}}{{Gustafsson} et~al.}{2008}]{Marcs}
{Gustafsson} B.,  {Edvardsson} B.,  {Eriksson} K.,  {J{\o}rgensen} U.~G.,
  {Nordlund} {\AA}.,   {Plez} B.,  2008, \mn@doi [\aap]
  {10.1051/0004-6361:200809724}, \href
  {http://adsabs.harvard.edu/abs/2008A%26A...486..951G} {486, 951}

\bibitem[\protect\citeauthoryear{{Hawkins} et~al.,}{{Hawkins}
  et~al.}{2020}]{Hawkins:2020aa}
{Hawkins} K.,  et~al., 2020, \mn@doi [\mnras] {10.1093/mnras/stz3132}, \href
  {https://ui.adsabs.harvard.edu/abs/2020MNRAS.492.1164H} {492, 1164}

\bibitem[\protect\citeauthoryear{{Heiter}, {Jofr{\'e}}, {Gustafsson}, {Korn},
  {Soubiran}  \& {Th{\'e}venin}}{{Heiter} et~al.}{2015}]{Heiter:2015aa}
{Heiter} U.,  {Jofr{\'e}} P.,  {Gustafsson} B.,  {Korn} A.~J.,  {Soubiran} C.,
   {Th{\'e}venin} F.,  2015, \mn@doi [\aap] {10.1051/0004-6361/201526319},
  \href {https://ui.adsabs.harvard.edu/abs/2015A&A...582A..49H} {582, A49}

\bibitem[\protect\citeauthoryear{{Hoppe}, {Bergemann}, {Bitsch}  \&
  {Serenelli}}{{Hoppe} et~al.}{2020}]{Hoppe:2020aa}
{Hoppe} R.,  {Bergemann} M.,  {Bitsch} B.,   {Serenelli} A.,  2020, \mn@doi
  [\aap] {10.1051/0004-6361/201936932}, \href
  {https://ui.adsabs.harvard.edu/abs/2020A&A...641A..73H} {641, A73}

\bibitem[\protect\citeauthoryear{{Huber} et~al.,}{{Huber}
  et~al.}{2012}]{huber:2012aa}
{Huber} D.,  et~al., 2012, \mn@doi [\apj] {10.1088/0004-637X/760/1/32}, \href
  {https://ui.adsabs.harvard.edu/abs/2012ApJ...760...32H} {760, 32}

\bibitem[\protect\citeauthoryear{{Izidoro}, {Bitsch}, {Raymond}, {Johansen},
  {Morbidelli}, {Lambrechts}  \& {Jacobson}}{{Izidoro}
  et~al.}{2021}]{Izidoro:2021aa}
{Izidoro} A.,  {Bitsch} B.,  {Raymond} S.~N.,  {Johansen} A.,  {Morbidelli} A.,
   {Lambrechts} M.,   {Jacobson} S.~A.,  2021, \mn@doi [\aap]
  {10.1051/0004-6361/201935336}, \href
  {https://ui.adsabs.harvard.edu/abs/2021A&A...650A.152I} {650, A152}

\bibitem[\protect\citeauthoryear{{Jermyn} et~al.,}{{Jermyn}
  et~al.}{2022}]{MESAVI}
{Jermyn} A.~S.,  et~al., 2022, \mn@doi [arXiv e-prints]
  {10.48550/arXiv.2208.03651}, \href
  {https://ui.adsabs.harvard.edu/abs/2022arXiv220803651J} {p. arXiv:2208.03651}

\bibitem[\protect\citeauthoryear{{Jofr{\'e}} et~al.,}{{Jofr{\'e}}
  et~al.}{2014}]{Jofre:2014aa}
{Jofr{\'e}} P.,  et~al., 2014, \mn@doi [\aap] {10.1051/0004-6361/201322440},
  \href {https://ui.adsabs.harvard.edu/abs/2014A&A...564A.133J} {564, A133}

\bibitem[\protect\citeauthoryear{{Jofr{\'e}} et~al.,}{{Jofr{\'e}}
  et~al.}{2015}]{Jofre:2015aa}
{Jofr{\'e}} P.,  et~al., 2015, \mn@doi [\aap] {10.1051/0004-6361/201526604},
  \href {https://ui.adsabs.harvard.edu/abs/2015A&A...582A..81J} {582, A81}

\bibitem[\protect\citeauthoryear{{Jofr{\'e}} et~al.,}{{Jofr{\'e}}
  et~al.}{2017}]{Jofre:2017aa}
{Jofr{\'e}} P.,  et~al., 2017, \mn@doi [\aap] {10.1051/0004-6361/201629833},
  \href {https://ui.adsabs.harvard.edu/abs/2017A&A...601A..38J} {601, A38}

\bibitem[\protect\citeauthoryear{{Johnson}, {Aller}, {Howard}  \&
  {Crepp}}{{Johnson} et~al.}{2010}]{Johnson:2010ab}
{Johnson} J.~A.,  {Aller} K.~M.,  {Howard} A.~W.,   {Crepp} J.~R.,  2010,
  \mn@doi [\pasp] {10.1086/655775}, \href
  {https://ui.adsabs.harvard.edu/abs/2010PASP..122..905J} {122, 905}

\bibitem[\protect\citeauthoryear{{Kamdar}, {Conroy}, {Ting}, {Bonaca}, {Smith}
  \& {Brown}}{{Kamdar} et~al.}{2019}]{Kamdar:2019aa}
{Kamdar} H.,  {Conroy} C.,  {Ting} Y.-S.,  {Bonaca} A.,  {Smith} M.~C.,
  {Brown} A. G.~A.,  2019, \mn@doi [\apjl] {10.3847/2041-8213/ab4997}, \href
  {https://ui.adsabs.harvard.edu/abs/2019ApJ...884L..42K} {884, L42}

\bibitem[\protect\citeauthoryear{{Kelson}}{{Kelson}}{2003}]{Kelson:2003aa}
{Kelson} D.~D.,  2003, \mn@doi [\pasp] {10.1086/375502}, \href
  {https://ui.adsabs.harvard.edu/abs/2003PASP..115..688K} {115, 688}

\bibitem[\protect\citeauthoryear{{Kollmeier} et~al.,}{{Kollmeier}
  et~al.}{2017}]{Kollmeier:2017aa}
{Kollmeier} J.~A.,  et~al., 2017, \mn@doi [arXiv e-prints]
  {10.48550/arXiv.1711.03234}, \href
  {https://ui.adsabs.harvard.edu/abs/2017arXiv171103234K} {p. arXiv:1711.03234}

\bibitem[\protect\citeauthoryear{{Korn}, {Grundahl}, {Richard}, {Mashonkina},
  {Barklem}, {Collet}, {Gustafsson}  \& {Piskunov}}{{Korn}
  et~al.}{2007}]{Korn:2007aa}
{Korn} A.~J.,  {Grundahl} F.,  {Richard} O.,  {Mashonkina} L.,  {Barklem}
  P.~S.,  {Collet} R.,  {Gustafsson} B.,   {Piskunov} N.,  2007, \mn@doi [\apj]
  {10.1086/523098}, \href
  {https://ui.adsabs.harvard.edu/abs/2007ApJ...671..402K} {671, 402}

\bibitem[\protect\citeauthoryear{{Krumholz} \& {Ting}}{{Krumholz} \&
  {Ting}}{2018}]{Krumholz:2018aa}
{Krumholz} M.~R.,  {Ting} Y.-S.,  2018, \mn@doi [\mnras]
  {10.1093/mnras/stx3286}, \href
  {https://ui.adsabs.harvard.edu/abs/2018MNRAS.475.2236K} {475, 2236}

\bibitem[\protect\citeauthoryear{{Liu}, {Asplund}, {Ramirez}, {Yong}  \&
  {Melendez}}{{Liu} et~al.}{2014}]{Liu:2014aa}
{Liu} F.,  {Asplund} M.,  {Ramirez} I.,  {Yong} D.,   {Melendez} J.,  2014,
  \mn@doi [\mnras] {10.1093/mnrasl/slu055}, \href
  {https://ui.adsabs.harvard.edu/abs/2014MNRAS.442L..51L} {442, L51}

\bibitem[\protect\citeauthoryear{{Liu}, {Yong}, {Asplund}, {Ram{\'\i}rez}  \&
  {Mel{\'e}ndez}}{{Liu} et~al.}{2016a}]{Liu:2016ab}
{Liu} F.,  {Yong} D.,  {Asplund} M.,  {Ram{\'\i}rez} I.,   {Mel{\'e}ndez} J.,
  2016a, \mn@doi [\mnras] {10.1093/mnras/stw247}, \href
  {https://ui.adsabs.harvard.edu/abs/2016MNRAS.457.3934L} {457, 3934}

\bibitem[\protect\citeauthoryear{{Liu}, {Asplund}, {Yong}, {Mel{\'e}ndez},
  {Ram{\'\i}rez}, {Karakas}, {Carlos}  \& {Marino}}{{Liu}
  et~al.}{2016b}]{Liu:2016aa}
{Liu} F.,  {Asplund} M.,  {Yong} D.,  {Mel{\'e}ndez} J.,  {Ram{\'\i}rez} I.,
  {Karakas} A.~I.,  {Carlos} M.,   {Marino} A.~F.,  2016b, \mn@doi [\mnras]
  {10.1093/mnras/stw2045}, \href
  {https://ui.adsabs.harvard.edu/abs/2016MNRAS.463..696L} {463, 696}

\bibitem[\protect\citeauthoryear{{Liu}, {Yong}, {Asplund}, {Feltzing},
  {Mustill}, {Mel{\'e}ndez}, {Ram{\'\i}rez}  \& {Lin}}{{Liu}
  et~al.}{2018}]{Liu:2018aa}
{Liu} F.,  {Yong} D.,  {Asplund} M.,  {Feltzing} S.,  {Mustill} A.~J.,
  {Mel{\'e}ndez} J.,  {Ram{\'\i}rez} I.,   {Lin} J.,  2018, \mn@doi [\aap]
  {10.1051/0004-6361/201832701}, \href
  {https://ui.adsabs.harvard.edu/abs/2018A&A...614A.138L} {614, A138}

\bibitem[\protect\citeauthoryear{{Liu}, {Asplund}, {Yong}, {Feltzing},
  {Dotter}, {Mel{\'e}ndez}  \& {Ram{\'\i}rez}}{{Liu} et~al.}{2019}]{Liu:2019aa}
{Liu} F.,  {Asplund} M.,  {Yong} D.,  {Feltzing} S.,  {Dotter} A.,
  {Mel{\'e}ndez} J.,   {Ram{\'\i}rez} I.,  2019, \mn@doi [\aap]
  {10.1051/0004-6361/201935306}, \href
  {https://ui.adsabs.harvard.edu/abs/2019A&A...627A.117L} {627, A117}

\bibitem[\protect\citeauthoryear{{Liu}, {Yong}, {Asplund}, {Wang}, {Spina},
  {Acu{\~n}a}, {Mel{\'e}ndez}  \& {Ram{\'\i}rez}}{{Liu}
  et~al.}{2020}]{Liu:2020aa}
{Liu} F.,  {Yong} D.,  {Asplund} M.,  {Wang} H.~S.,  {Spina} L.,  {Acu{\~n}a}
  L.,  {Mel{\'e}ndez} J.,   {Ram{\'\i}rez} I.,  2020, \mn@doi [\mnras]
  {10.1093/mnras/staa1420}, \href
  {https://ui.adsabs.harvard.edu/abs/2020MNRAS.495.3961L} {495, 3961}

\bibitem[\protect\citeauthoryear{{Liu}, {Bitsch}, {Asplund}, {Liu}, {Murphy},
  {Yong}, {Ting}  \& {Feltzing}}{{Liu} et~al.}{2021}]{Liu:2021aa}
{Liu} F.,  {Bitsch} B.,  {Asplund} M.,  {Liu} B.-B.,  {Murphy} M.~T.,  {Yong}
  D.,  {Ting} Y.-S.,   {Feltzing} S.,  2021, \mn@doi [\mnras]
  {10.1093/mnras/stab2471}, \href
  {https://ui.adsabs.harvard.edu/abs/2021MNRAS.508.1227L} {508, 1227}

\bibitem[\protect\citeauthoryear{{Majewski} et~al.,}{{Majewski}
  et~al.}{2017}]{Majewski:2017aa}
{Majewski} S.~R.,  et~al., 2017, \mn@doi [\aj] {10.3847/1538-3881/aa784d},
  \href {https://ui.adsabs.harvard.edu/abs/2017AJ....154...94M} {154, 94}

\bibitem[\protect\citeauthoryear{{Mel{\'e}ndez}, {Asplund}, {Gustafsson}  \&
  {Yong}}{{Mel{\'e}ndez} et~al.}{2009}]{Melendez:2009ab}
{Mel{\'e}ndez} J.,  {Asplund} M.,  {Gustafsson} B.,   {Yong} D.,  2009, \mn@doi
  [\apjl] {10.1088/0004-637X/704/1/L66}, \href
  {https://ui.adsabs.harvard.edu/abs/2009ApJ...704L..66M} {704, L66}

\bibitem[\protect\citeauthoryear{{Mel{\'e}ndez} et~al.,}{{Mel{\'e}ndez}
  et~al.}{2012}]{Melendez:2012aa}
{Mel{\'e}ndez} J.,  et~al., 2012, \mn@doi [\aap] {10.1051/0004-6361/201117222},
  \href {https://ui.adsabs.harvard.edu/abs/2012A&A...543A..29M} {543, A29}

\bibitem[\protect\citeauthoryear{{Mulders}, {Pascucci}, {Apai}  \&
  {Ciesla}}{{Mulders} et~al.}{2018}]{Mulders:2018aa}
{Mulders} G.~D.,  {Pascucci} I.,  {Apai} D.,   {Ciesla} F.~J.,  2018, \mn@doi
  [\aj] {10.3847/1538-3881/aac5ea}, \href
  {https://ui.adsabs.harvard.edu/abs/2018AJ....156...24M} {156, 24}

\bibitem[\protect\citeauthoryear{{Nagar}, {Spina}  \& {Karakas}}{{Nagar}
  et~al.}{2020}]{nagar:2020aa}
{Nagar} T.,  {Spina} L.,   {Karakas} A.~I.,  2020, \mn@doi [\apjl]
  {10.3847/2041-8213/ab5dc6}, \href
  {https://ui.adsabs.harvard.edu/abs/2020ApJ...888L...9N} {888, L9}

\bibitem[\protect\citeauthoryear{{Nelson}, {Ting}, {Hawkins}, {Ji}, {Kamdar}
  \& {El-Badry}}{{Nelson} et~al.}{2021}]{Nelson:2021aa}
{Nelson} T.,  {Ting} Y.-S.,  {Hawkins} K.,  {Ji} A.,  {Kamdar} H.,   {El-Badry}
  K.,  2021, \mn@doi [\apj] {10.3847/1538-4357/ac14be}, \href
  {https://ui.adsabs.harvard.edu/abs/2021ApJ...921..118N} {921, 118}

\bibitem[\protect\citeauthoryear{{Nissen}}{{Nissen}}{2015}]{Nissen:2015aa}
{Nissen} P.~E.,  2015, \mn@doi [\aap] {10.1051/0004-6361/201526269}, \href
  {https://ui.adsabs.harvard.edu/abs/2015A&A...579A..52N} {579, A52}

\bibitem[\protect\citeauthoryear{{Nissen} \& {Gustafsson}}{{Nissen} \&
  {Gustafsson}}{2018}]{Nissen:2018aa}
{Nissen} P.~E.,  {Gustafsson} B.,  2018, \mn@doi [\aapr]
  {10.1007/s00159-018-0111-3}, \href
  {https://ui.adsabs.harvard.edu/abs/2018A&ARv..26....6N} {26, 6}

\bibitem[\protect\citeauthoryear{{Nordlander}, {Korn}, {Richard}  \&
  {Lind}}{{Nordlander} et~al.}{2012}]{Nordlander:2012aa}
{Nordlander} T.,  {Korn} A.~J.,  {Richard} O.,   {Lind} K.,  2012, \mn@doi
  [\apj] {10.1088/0004-637X/753/1/48}, \href
  {https://ui.adsabs.harvard.edu/abs/2012ApJ...753...48N} {753, 48}

\bibitem[\protect\citeauthoryear{{Oh}, {Price-Whelan}, {Brewer}, {Hogg},
  {Spergel}  \& {Myles}}{{Oh} et~al.}{2018}]{Oh:2018aa}
{Oh} S.,  {Price-Whelan} A.~M.,  {Brewer} J.~M.,  {Hogg} D.~W.,  {Spergel}
  D.~N.,   {Myles} J.,  2018, \mn@doi [\apj] {10.3847/1538-4357/aaab4d}, \href
  {https://ui.adsabs.harvard.edu/abs/2018ApJ...854..138O} {854, 138}

\bibitem[\protect\citeauthoryear{{Paxton}, {Bildsten}, {Dotter}, {Herwig},
  {Lesaffre}  \& {Timmes}}{{Paxton} et~al.}{2011a}]{MESAI}
{Paxton} B.,  {Bildsten} L.,  {Dotter} A.,  {Herwig} F.,  {Lesaffre} P.,
  {Timmes} F.,  2011a, \mn@doi [\apjs] {10.1088/0067-0049/192/1/3}, \href
  {http://adsabs.harvard.edu/abs/2011ApJS..192....3P} {192, 3}

\bibitem[\protect\citeauthoryear{{Paxton}, {Bildsten}, {Dotter}, {Herwig},
  {Lesaffre}  \& {Timmes}}{{Paxton} et~al.}{2011b}]{paxton:2011aa}
{Paxton} B.,  {Bildsten} L.,  {Dotter} A.,  {Herwig} F.,  {Lesaffre} P.,
  {Timmes} F.,  2011b, \mn@doi [\apjs] {10.1088/0067-0049/192/1/3}, \href
  {https://ui.adsabs.harvard.edu/abs/2011ApJS..192....3P} {192, 3}

\bibitem[\protect\citeauthoryear{{Paxton} et~al.,}{{Paxton}
  et~al.}{2013}]{MESAII}
{Paxton} B.,  et~al., 2013, \mn@doi [\apjs] {10.1088/0067-0049/208/1/4}, \href
  {http://adsabs.harvard.edu/abs/2013ApJS..208....4P} {208, 4}

\bibitem[\protect\citeauthoryear{{Paxton} et~al.,}{{Paxton}
  et~al.}{2015}]{MESAIII}
{Paxton} B.,  et~al., 2015, \mn@doi [\apjs] {10.1088/0067-0049/220/1/15}, \href
  {http://adsabs.harvard.edu/abs/2015ApJS..220...15P} {220, 15}

\bibitem[\protect\citeauthoryear{{Paxton} et~al.,}{{Paxton}
  et~al.}{2018}]{MESAIV}
{Paxton} B.,  et~al., 2018, \mn@doi [\apjs] {10.3847/1538-4365/aaa5a8}, \href
  {http://adsabs.harvard.edu/abs/2018ApJS..234...34P} {234, 34}

\bibitem[\protect\citeauthoryear{{Paxton} et~al.,}{{Paxton}
  et~al.}{2019}]{MESAV}
{Paxton} B.,  et~al., 2019, \mn@doi [\apjs] {10.3847/1538-4365/ab2241}, \href
  {https://ui.adsabs.harvard.edu/abs/2019ApJS..243...10P} {243, 10}

\bibitem[\protect\citeauthoryear{{Pinsonneault}, {DePoy}  \&
  {Coffee}}{{Pinsonneault} et~al.}{2001}]{Pinsonneault:2001aa}
{Pinsonneault} M.~H.,  {DePoy} D.~L.,   {Coffee} M.,  2001, \mn@doi [\apjl]
  {10.1086/323531}, \href
  {https://ui.adsabs.harvard.edu/abs/2001ApJ...556L..59P} {556, L59}

\bibitem[\protect\citeauthoryear{{Rains}, {Ireland}, {White}, {Casagrande}  \&
  {Karovicova}}{{Rains} et~al.}{2020}]{rains:2020aa}
{Rains} A.~D.,  {Ireland} M.~J.,  {White} T.~R.,  {Casagrande} L.,
  {Karovicova} I.,  2020, \mn@doi [\mnras] {10.1093/mnras/staa282}, \href
  {https://ui.adsabs.harvard.edu/abs/2020MNRAS.493.2377R} {493, 2377}

\bibitem[\protect\citeauthoryear{{Ram{\'{\i}}rez} \&
  {Mel{\'e}ndez}}{{Ram{\'{\i}}rez} \& {Mel{\'e}ndez}}{2005}]{Ramirez:2005ab}
{Ram{\'{\i}}rez} I.,  {Mel{\'e}ndez} J.,  2005, \mn@doi [\apj]
  {10.1086/430102}, \href {http://adsabs.harvard.edu/abs/2005ApJ...626..465R}
  {626, 465}

\bibitem[\protect\citeauthoryear{{Ram{\'\i}rez} et~al.,}{{Ram{\'\i}rez}
  et~al.}{2014}]{ramirez:2014aa}
{Ram{\'\i}rez} I.,  et~al., 2014, \mn@doi [\aap] {10.1051/0004-6361/201424244},
  \href {https://ui.adsabs.harvard.edu/abs/2014A&A...572A..48R} {572, A48}

\bibitem[\protect\citeauthoryear{{Ram{\'\i}rez} et~al.,}{{Ram{\'\i}rez}
  et~al.}{2015}]{ramirez:2015aa}
{Ram{\'\i}rez} I.,  et~al., 2015, \mn@doi [\apj] {10.1088/0004-637X/808/1/13},
  \href {https://ui.adsabs.harvard.edu/abs/2015ApJ...808...13R} {808, 13}

\bibitem[\protect\citeauthoryear{{Ram{\'\i}rez}, {Khanal}, {Lichon},
  {Chanam{\'e}}, {Endl}, {Mel{\'e}ndez}  \& {Lambert}}{{Ram{\'\i}rez}
  et~al.}{2019}]{Ramirez:2019aa}
{Ram{\'\i}rez} I.,  {Khanal} S.,  {Lichon} S.~J.,  {Chanam{\'e}} J.,  {Endl}
  M.,  {Mel{\'e}ndez} J.,   {Lambert} D.~L.,  2019, \mn@doi [\mnras]
  {10.1093/mnras/stz2709}, \href
  {https://ui.adsabs.harvard.edu/abs/2019MNRAS.490.2448R} {490, 2448}

\bibitem[\protect\citeauthoryear{{Rosenthal} et~al.,}{{Rosenthal}
  et~al.}{2022}]{Rosenthal:2022aa}
{Rosenthal} L.~J.,  et~al., 2022, \mn@doi [\apjs] {10.3847/1538-4365/ac7230},
  \href {https://ui.adsabs.harvard.edu/abs/2022ApJS..262....1R} {262, 1}

\bibitem[\protect\citeauthoryear{{Saffe}, {Jofr{\'e}}, {Martioli}, {Flores},
  {Petrucci}  \& {Jaque Arancibia}}{{Saffe} et~al.}{2017}]{Saffe:2017aa}
{Saffe} C.,  {Jofr{\'e}} E.,  {Martioli} E.,  {Flores} M.,  {Petrucci} R.,
  {Jaque Arancibia} M.,  2017, \mn@doi [\aap] {10.1051/0004-6361/201731430},
  \href {https://ui.adsabs.harvard.edu/abs/2017A&A...604L...4S} {604, L4}

\bibitem[\protect\citeauthoryear{{Sneden}}{{Sneden}}{1973}]{Sneden:1973aa}
{Sneden} C.,  1973, \mn@doi [\apj] {10.1086/152374}, \href
  {http://adsabs.harvard.edu/abs/1973ApJ...184..839S} {184, 839}

\bibitem[\protect\citeauthoryear{{Sobeck} et~al.,}{{Sobeck}
  et~al.}{2011}]{Sobeck:2011aa}
{Sobeck} J.~S.,  et~al., 2011, \mn@doi [\aj] {10.1088/0004-6256/141/6/175},
  \href {http://adsabs.harvard.edu/abs/2011AJ....141..175S} {141, 175}

\bibitem[\protect\citeauthoryear{{Souto} et~al.,}{{Souto}
  et~al.}{2018}]{Souto:2018aa}
{Souto} D.,  et~al., 2018, \mn@doi [\apj] {10.3847/1538-4357/aab612}, \href
  {https://ui.adsabs.harvard.edu/abs/2018ApJ...857...14S} {857, 14}

\bibitem[\protect\citeauthoryear{{Souto} et~al.,}{{Souto}
  et~al.}{2019}]{Souto:2019aa}
{Souto} D.,  et~al., 2019, \mn@doi [\apj] {10.3847/1538-4357/ab0b43}, \href
  {https://ui.adsabs.harvard.edu/abs/2019ApJ...874...97S} {874, 97}

\bibitem[\protect\citeauthoryear{{Spina} et~al.,}{{Spina}
  et~al.}{2018}]{spina:2018aa}
{Spina} L.,  et~al., 2018, \mn@doi [\mnras] {10.1093/mnras/stx2938}, \href
  {https://ui.adsabs.harvard.edu/abs/2018MNRAS.474.2580S} {474, 2580}

\bibitem[\protect\citeauthoryear{{Spina}, {Sharma}, {Mel{\'e}ndez}, {Bedell},
  {Casey}, {Carlos}, {Franciosini}  \& {Vallenari}}{{Spina}
  et~al.}{2021}]{Spina:2021aa}
{Spina} L.,  {Sharma} P.,  {Mel{\'e}ndez} J.,  {Bedell} M.,  {Casey} A.~R.,
  {Carlos} M.,  {Franciosini} E.,   {Vallenari} A.,  2021, \mn@doi [Nature
  Astronomy] {10.1038/s41550-021-01451-8}, \href
  {https://ui.adsabs.harvard.edu/abs/2021NatAs...5.1163S} {5, 1163}

\bibitem[\protect\citeauthoryear{{Tayar}, {Claytor}, {Huber}  \& {van
  Saders}}{{Tayar} et~al.}{2022}]{tayar:2022aa}
{Tayar} J.,  {Claytor} Z.~R.,  {Huber} D.,   {van Saders} J.,  2022, \mn@doi
  [\apj] {10.3847/1538-4357/ac4bbc}, \href
  {https://ui.adsabs.harvard.edu/abs/2022ApJ...927...31T} {927, 31}

\bibitem[\protect\citeauthoryear{{Tucci Maia}, {Mel{\'e}ndez}  \&
  {Ram{\'\i}rez}}{{Tucci Maia} et~al.}{2014}]{tuccimaia:2014aa}
{Tucci Maia} M.,  {Mel{\'e}ndez} J.,   {Ram{\'\i}rez} I.,  2014, \mn@doi
  [\apjl] {10.1088/2041-8205/790/2/L25}, \href
  {https://ui.adsabs.harvard.edu/abs/2014ApJ...790L..25T} {790, L25}

\bibitem[\protect\citeauthoryear{{Ulrich}}{{Ulrich}}{1972}]{Ulrich:1972aa}
{Ulrich} R.~K.,  1972, \mn@doi [\apj] {10.1086/151336}, \href
  {https://ui.adsabs.harvard.edu/abs/1972ApJ...172..165U} {172, 165}

\bibitem[\protect\citeauthoryear{{Vogt} et~al.,}{{Vogt}
  et~al.}{1994}]{Vogt:1994aa}
{Vogt} S.~S.,  et~al., 1994, in {Crawford} D.~L.,  {Craine} E.~R.,  eds,
  Society of Photo-Optical Instrumentation Engineers (SPIE) Conference Series
  Vol. 2198, Instrumentation in Astronomy VIII. p.~362,
  \mn@doi{10.1117/12.176725}

\bibitem[\protect\citeauthoryear{{de Jong} et~al.,}{{de Jong}
  et~al.}{2019}]{deJong:2019aa}
{de Jong} R.~S.,  et~al., 2019, \mn@doi [The Messenger]
  {10.18727/0722-6691/5117}, \href
  {https://ui.adsabs.harvard.edu/abs/2019Msngr.175....3D} {175, 3}

\makeatother
\end{thebibliography}

\input{ms.bbl}




%


\bsp	
\label{lastpage}
\end{document}